\documentclass[twocolumn,iop]{./emulateapj}

\usepackage{graphics}
 \usepackage{epstopdf}

\usepackage{amssymb}
\usepackage{amsmath}
\usepackage{natbib}
\usepackage{lineno}

\slugcomment{Manuscript published in Astrophysical Journal doi:10.1088/0004-637X/751/1/32}

\makeatletter

\def\compoundrel#1\over#2{\mathpalette\compoundreL{{#1}\over{#2}}}
\def\compoundreL#1#2{\compoundREL#1#2}
\def\compoundREL#1#2\over#3{\mathrel
      {\vcenter{\hbox{$\m@th\buildrel{#1#2}\over{#1#3}$}}}}
\makeatother

\def\aj{AJ}
\def\apj{ApJ}
\def\apjl{ApJL}
\def\nat{Nature}
\def\icarus{Icarus}
\def\araa{Ann. Rev. Astron. \& Astrophys.}
\def\areps{Ann. Rev. Earth Planet Sci.}
\def\mnras{MNRAS}
\def\aap{A\&A}

\def\ssr{Space Sci. Rev.}
\def\apjs{ApJS}
\def\chemie{Chemie der Erde}
\def\aspcs{ASP Conf. Series}
\def\epsl{Earth Planet. Sci. Lett.}
\def\lpsc{Lunar Planet. Sci. Conf.}
\def\baas{BAAS}
\def\maps{Met. Planet. Sci.}

\begin{document}

\title{Collisions between Gravity-Dominated Bodies: \\ 
2. The Diversity of Impact Outcomes during the End Stage of Planet Formation \\}

\author{Sarah T. Stewart$^1$ and Zo\"e M. Leinhardt$^2$}

\affil{$^1$Department of Earth and Planetary Sciences, Harvard
  University, 20 Oxford Street, Cambridge, MA 02138, U.S.A. \\
$^2$School of Physics, University of Bristol, H. H. Wills Physics
Laboratory, Tyndall Avenue, BS8 1TL, U.K. }
\email{sstewart@eps.harvard.edu}
\email{zoe.leinhardt@bristol.ac.uk}

\begin{abstract}
Numerical simulations of the stochastic end stage of planet formation typically begin with a population of embryos and planetesimals that grow into planets by merging.  We analyzed the impact parameters of collisions leading to the growth of terrestrial planets from recent $N$-body simulations that assumed perfect merging and calculated more realistic outcomes using a new analytic collision physics model. We find that collision outcomes are diverse and span all possible regimes: hit-and-run, merging, partial accretion, partial erosion, and catastrophic disruption.  The primary outcomes of giant impacts between planetary embryos are approximately evenly split between partial accretion, graze-and-merge, and hit-and-run events.  To explore the cumulative effects of more realistic collision outcomes, we modeled the growth of individual planets with a Monte Carlo technique using the distribution of impact parameters from $N$-body simulations.  We find that fewer planets reached masses $>0.7 M_{\rm Earth}$ using the collision physics model compared to simulations that assumed every collision results in perfect merging. For final planets with masses $>0.7 M_{\rm Earth}$, 60\% are enriched in their core-to-mantle mass fraction by $>10$\% compared to the initial embryo composition.  Fragmentation during planet formation produces significant debris ($\sim15$\% of the final mass) and occurs primarily by erosion of the smaller body in partial accretion and hit-and-run events.  In partial accretion events, the target body grows by preferentially accreting the iron core of the projectile and the escaping fragments are derived primarily from the silicate mantles of both bodies. Thus, the bulk composition of a planet can evolve via stochastic giant impacts.

\end{abstract}


\lefthead{The Astrophysical Journal}
\righthead{Stewart \& Leinhardt: Diverse Giant Impact Outcomes}

\normalsize

\section{Introduction}
The end stage of terrestrial planet formation is defined by the final
competition between planetary embryos as they grow via collisions into
a final set of stable planets \citep[e.g.,][]{Chambers:2004}. The last
giant impact is invoked to explain major differences between the solid
planets in our Solar System. The large core mass fraction for Mercury
could be the result of a single mantle-stripping impact event
\citep{Benz:2007}. A planet's spin orientation and period is strongly
influenced by the last giant impact \citep{Lissauer:1993}. Giant
collisions are also invoked to explain the formation of Earth's moon
\citep{Canup:2001}, the Pluto system \citep{Canup:2005}, and the
Haumea system \citep{Leinhardt:2010}.

The growth of planetary embryos into planets is commonly modeled using
$N$-body techniques in order to capture the details of the multiple
gravitational interactions. Because typical encounter velocities
between embryos are in the range of 1 to 3 mutual escape velocities
\citep[e.g.,][]{Agnor:1999}, almost all $N$-body simulations have
assumed that the two colliding embryos merge into a single body with
the same total mass. However, hydrocode simulations of oblique
collisions between embryos found that inelastic collisions, known as
hit-and-run (H\&R) events \citep{Agnor:2004, Asphaug:2006}, are a
common outcome at speeds just above the mutual escape
velocity. \citet{Asphaug:2010} estimated that about half of giant
impacts are hit-and-run events. The physics of giant impacts is
sufficiently complicated that the impact parameters demarcating the
transition between merging and hit-and-run outcomes was defined only
recently by an empirical analysis of over a thousand hydrocode impact
simulations \citep{Kokubo:2010,Genda:2012}.

Collision outcomes other than merging and hit-and-run are possible
during planet formation. The range of outcomes include partial
accretion, partial erosion, and catastrophically disruptive
events. Some planet formation simulations, focusing on specific stages
of planet growth, have directly modeled each encounter between two
bodies in order to capture the details of the collision physics
\citep{Genda:2011,Leinhardt:2005a,Leinhardt:2009a}, but this approach
is too computationally expensive to be widely adopted. To date,
planet formation studies have not had access to a robust collision
physics model that could be implemented easily into $N$-body
simulations.

In a companion paper \citep{Leinhardt:2012}, we have developed a
comprehensive analytic model to predict the outcome of any collision
between gravity-dominated bodies. The model predicts the transitions
between collision regimes and the size and velocity distribution of
fragments. Collision outcomes depend on the projectile-to-target mass
ratio, impact angle, impact velocity, and two material parameters that
define the catastrophic disruption criteria. Catastrophic disruption
is defined by the specific impact energy needed to disperse half the
total mass.  For collisions that result in partial accretion or
erosion of differentiated bodies, we estimate the change in the
core-mantle mass ratio following the work of
\citet{Marcus:2009,Marcus:2010}. The new analytic model now allows
planet formation simulations to capture the full diversity of
collision outcomes.

To assess the effects of including more realistic collision physics in
planet formation studies, we analyzed the impact parameters from
recent $N$-body simulations of terrestrial planet formation that
assumed perfect merging and calculated the more realistic collision
outcomes \citep{Obrien:2006,Raymond:2009}. The results provide
insights into the most important collision regimes, quantify the
probabilities of different outcomes, and identify future challenges in
planet formation simulations.

Then, using the distribution of impact parameters from the $N$-body
simulations, we modeled individual planet growth with a Monte Carlo
technique in order to assess the cumulative effects of diverse
collision outcomes. We compare the distribution of final planet masses
from simulations with different assumptions about collision outcomes
and tracked changes in the core mass fraction. The collision physics
model allows for more detailed investigation of the evolution of the
bulk chemistry and for tighter isotopic constraints on terrestrial
planet formation studies.

\section{The Collision Outcome Model} \label{sec:collmodel}

\begin{figure*}[t]
\begin{center}
\includegraphics[width=35pc]{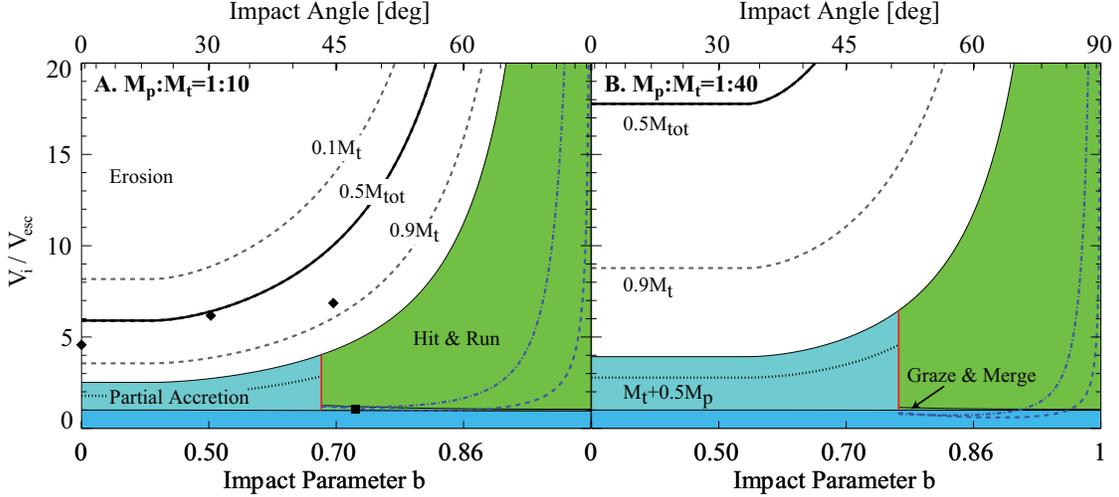}
\end{center}
\caption{Predicted collision outcome maps using the analytic model for
  strengthless planets (with material parameters $c^*=1.9$ and $\bar
  \mu=0.36$) for two projectile-to-target mass ratios of (A) $M_{\rm
    p}:M_{\rm t}=1:10$ and (B) 1:40.  Impact velocity is normalized by
  the mutual surface escape velocity; impact parameter is spaced
  according to probability. Colored regions denote perfect merging
  (dark blue), partial accretion to the target (medium blue), net
  erosion to the target (white), hit-and-run (green), and
  graze-and-merge (blue-green). Vertical red line at $b_{\rm crit}$
  denotes the transition between non-grazing and grazing events. Thick
  black curve -- critical velocity for catastrophic disruption; dashed
  grey curves -- 10\% and 90\% of target mass in largest remnant;
  dotted black curve -- 50\% of projectile accreted; dot-dashed blue
  curve -- catastrophic disruption of the projectile in a hit-and-run;
  dashed blue curve -- onset of erosion of projectile in a
  hit-and-run. Symbols denote proposed giant impact events:
  $\blacklozenge$ -- Mercury \citep{Benz:2007}; $\blacksquare$ --
  Earth-Moon \citep{Canup:2004}.\\}
 \label{fig:collmap}
 \end{figure*}

The outcome of a collision in the gravity regime was predicted using
the new analytic collision physics model from \citet{Leinhardt:2012},
hereafter LS12. The model is summarized here; refer to LS12 for the
full details and limitations. In the gravity regime, erosive outcomes
are primarily controlled by overcoming gravitational forces rather
than overcoming material strength; bodies larger than about 100 to
1000~m in size are dominated by gravity
\citep[e.g.,][]{Melosh:1997}. The model predicts the collision outcome
as a function of projectile-to-target mass ratio ($\gamma=M_{\rm
  p}/M_{\rm t}$), impact angle ($\theta$), and impact velocity
($V_i$).  By convention, the projectile is less massive than the
target ($M_{\rm p} \le M_{\rm t}$). The following equations were used
to generate the boundaries and contours on the example collision
outcome maps shown in Figure~\ref{fig:collmap}\footnote{A computer
  code to calculate individual collision outcomes and generate
  collision outcome maps is available from the authors.}.

Partial accretion and erosive outcomes are defined by the catastrophic
disruption criteria ($Q^*_{RD}$), which is the specific impact energy
required to disrupt and gravitationally disperse half the total mass.
For a particular collision, the total specific impact energy in the
center of mass frame is given by
\begin{equation}
Q_R = \frac{\mu V_i^2 }{2 M_{\rm tot}}, \label{eqn:qr}
\end{equation}
where $M_{\rm tot}=M_{\rm p}+M_{\rm t}$ and $\mu=M_{\rm p}M_{\rm
  t}/M_{\rm tot}$ is the reduced mass. For bodies in the gravity
regime, the head-on disruption criteria increases with increasing mass
and follows the general form derived by \citet{Housen:1990}:
\begin{equation} \label{eqn:qstarred} 
Q^*_{RD}= q_g \left (\rho_1 G \right )^{3 \bar \mu/2} R_{C1}^{3 \bar \mu} V^{*(2-3\bar \mu)},
\end{equation}
where $R_{C1}$ is the spherical radius of the combined projectile and
target masses at a density of $\rho_1 \equiv 1000$~kg~m$^{-3}$ and $G$
is the gravitational constant. At the critical impact velocity
$(V_i=V^*)$, the specific impact energy is equal to the catastrophic
disruption threshold $(Q_R=Q^*_{RD})$ for the particular impact
scenario ($R_{C1}$ and mass ratio). Instead of target radius, the
variable $R_{C1}$ was introduced by \citet{Stewart:2009} in order to
generalize the dependence of $Q^*_{RD}$ on size for different density
bodies and for different projectile-to-target mass ratios.  The
coefficient $q_g$ is of order 1 and defined below. $\bar \mu$ is the
velocity exponent in the point source coupling parameter from
\citet{Housen:1990}. LS12 fit a value of $\bar \mu=0.36\pm0.01$ using
the results from several numerical studies of disruption in the
gravity regime on a wide variety of materials (weak rock, ice, rubble
piles).

LS12 derived the disruption criteria for a reference case of a head-on
impact between two equal-mass bodies, which represents the minimum
energy scenario to disrupt a particular target. The critical impact
velocity for head-on equal mass collisions is given by
\begin{equation} \label{eqn:vequal}
V^{*}_{\gamma=1} = \left ( \frac{32 \pi c^*}{5} \right )^{1/2} (\rho_1 G)^{1/2} R_{C1},
\end{equation}
where $c^*$ is a nondimensional material parameter that is fitted to
simulation data. LS12 fit $c^*=1.9\pm0.3$ for fluid planets and
$c^*=5\pm2$ for various solid planetesimal analogs using the results
from several numerical studies.  $c^*$ is defined such that it
represents the value of the equal-mass head-on disruption criteria in
units of the specific gravitational binding energy of the combined
mass:
\begin{equation} \label{eqn:principalq}
  Q^*_{RD,\gamma=1}= c^* \frac{4}{5} \pi \rho_1 G R_{C1}^2.
\end{equation}
Equation \ref{eqn:principalq} is named the ``principal disruption
curve.''  Combining the previous equations gives the coefficient for
Equation~\ref{eqn:qstarred}:
\begin{equation}
q_g  =  \frac{1}{8} \left ( \frac{32 \pi c^*}{5} \right )^{3 \bar  \mu / 2}.
\end{equation}

The derivation of Equation \ref{eqn:qstarred} assumed that all of the
projectile's kinetic energy is deposited in the target. This
assumption is often invalid as a portion of the projectile misses the
target for many combinations of projectile-to-target mass ratio and
impact angle.  For an oblique impact, LS12 derived the mass fraction of
the projectile that geometrically overlaps with the target,
\begin{equation}
\alpha  = 
\begin{cases}
   1 & \text{if $R_{\rm t}> b (R_{\rm p} + R_{\rm t})+R_{\rm p}$,} \\
   \rho (\pi R_{\rm p} l^2 - \pi l^3 / 3) & \text{otherwise,} 
\end{cases}
\end{equation}
where $\rho$ is the density of the projectile, $b=\sin\theta$ is the
impact parameter ($b=0$ for head-on impacts) and $l/(2R_{\rm p})$ is
the fraction of the projectile diameter that overlaps the target. The
impact angle and overlapping mass fraction are defined at the point of
first contact assuming spherical bodies.  Using just the interacting
fraction the projectile, the angle-adjusted reduced mass is
\begin{equation}
  \mu_{\alpha}= \frac{\alpha M_{\rm p} M_{\rm t}}{\alpha M_{\rm p} +
    M_{\rm t}}. \label{eqn:mualpha}
\end{equation}

The disruption criteria and critical impact velocity for a specific
mass ratio and impact angle are calculated by adjustments to the
head-on equal-mass principal disruption curve. To achieve disruption
with a smaller projectile, the impact velocity rises to
\begin{equation} \label{eqn:vbar}
V^* = \left [ \frac{1}{4} \frac{(\gamma+1)^2}{\gamma} \right ]^{1/(3 \bar
  \mu)}  V^*_{\gamma=1},
\end{equation}
and, from Equation \ref{eqn:qstarred}, the catastrophic disruption
criteria increases by
\begin{eqnarray} \label{eqn:qmassratio}
Q^{*}_{RD} & = & Q^{*}_{RD,\gamma=1} \left ( \frac{V^*}{V^*_{\gamma=1}} \right )^{2-3\bar \mu}, \nonumber \\
       & = & Q^{*}_{RD,\gamma=1} \left [ \frac{1}{4} \frac{(\gamma+1)^2}{\gamma} \right ]^{2/(3 \bar \mu)-1} .
\end{eqnarray}
Next, accounting for the projectile interacting mass fraction for oblique impacts,
\begin{equation}
Q^{\prime *}_{RD}   = \left ( \frac{\mu}{\mu_{\alpha}} \right )^{2-3 \bar \mu /2} Q^{*}_{RD}, 
\end{equation}
where the prime notation indicates the value for an oblique
impact. Finally, from Equation \ref{eqn:qr}, the critical impact
velocity adjusted for both the mass ratio and impact angle is
\begin{equation}
  V^{\prime *} = \sqrt{ 2 Q^{\prime *}_{RD} M_{\rm tot} / \mu }.
\end{equation}
The equations for $Q^{\prime *}_{RD}$ and $V^{\prime *}$ are curves as
a function of $R_{C1}$, impact angle (within $\mu_{\alpha}$), mass
ratio (within $\mu$), and two material parameters, $c^*$ and $\bar
\mu$.

The specific collision outcome regime is determined using the mutual
escape velocity, disruption criteria, and impact parameter. When the
impact velocity is below the mutual escape velocity of the interacting
mass ($M^{\prime}_{\rm tot}=\alpha M_{\rm p}+M_{\rm t}$), the two
bodies are assumed to merge completely (perfect merging).  Above
$V^{\prime}_{\rm esc}=\sqrt{2 G M^{\prime}_{\rm tot} / (R_{\rm
    p}+R_{\rm t})}$, the collisions outcomes are divided into two
groups: grazing and non-grazing.

The transition between non-grazing and grazing outcomes is demarcated
by a critical impact parameter, $b_{\rm crit}$. Following
\citet{Asphaug:2010}, the center of the projectile is tangent to the
target at the critical impact parameter,
\begin{equation} \label{eqn:bcrit}
 b_{\rm crit} =\left( \frac{R}{R+r} \right).
\end{equation}
Non-grazing ($b<b_{\rm crit}$) collisions transition from perfect
merging to the disruption regime with increasing impact velocity. In
the disruption regime, the outcome may be partial accretion or partial
erosion depending on the mass of the largest remnant, $M_{\rm
  lr}$. The largest remnant mass is calculated using the impact energy
and the catastrophic disruption criteria. For collisions with
$0<Q_R/Q^{\prime *}_{RD}<1.8$, $M_{\rm lr}$ is proportional to impact
energy:
\begin{equation} \label{eqn:univlaw} 
 \frac{M_{\rm lr}}{M_{\rm tot}} = -0.5 \left ( \frac{Q_R}{Q^{\prime *}_{RD}} -1 \right ) + 0.5.
\end{equation}
Because a single line fit a wide variety of simulation results, we
named this linear relationship the ``universal law for the mass of
the largest remnant'' \citep{Stewart:2009}.  In the case of
super-catastrophic disruption, $Q_R/Q^{\prime *}_{RD} \ge 1.8$, the
largest remnant follows a power law,
\begin{equation} \label{eqn:mlrsuper}
\frac{M_{\rm lr}}{M_{\rm tot}} = \frac{0.1}{1.8^{\eta}} \left (
  \frac{Q_R}{Q^{\prime *}_{RD}} \right )^{\eta},
\end{equation}
where $\eta \sim -1.5$ based on laboratory disruption experiments, as
discussed in LS12.

Grazing ($b>b_{\rm crit}$) collisions transition from perfect merging
to a graze-and-merge regime, where the two bodies hit and separate
(cleanly or not) but are gravitationally bound and ultimately merge
\citep{Leinhardt:2010}.  The boundary between graze-and-merge and
hit-and-run, where the two bodies hit and then escape each other with
the target relatively intact, is difficult to define. From over a
thousand hydrocode collision simulations, \citet{Kokubo:2010} fit an
empirical relation for the velocity of transition to hit-and-run,
$V_{\rm hr}$,
\begin{equation} \label{eqn:vhr}
\frac{V_{\rm hr}}{V_{\rm esc}}=c_1 \zeta^2 (1-b)^{5/2} + c_2 \zeta^2 +c_3 (1-b)^{5/2}+c_4,
\end{equation}
where $\zeta=(M_{\rm t}-M_{\rm p})/M_{\rm tot}$, $c_1=2.43$,
$c_2=-0.0408$, $c_3=1.86$, and $c_4=1.08$ (see also
\citet{Genda:2012}).  We stress that the transition between
graze-and-merge and hit-and-run is likely to depend on material
properties of the planet. These coefficients are fit to smoothed
particle hydrodynamics simulations of fluid Earth-mass planets.

In $N$-body simulations, bodies are described by their mass, and radii
are usually estimated assuming a constant bulk density in order to
calculate the impact angle and mutual surface escape velocity, $V_{\rm
  esc}=\sqrt{2GM_{\rm tot}/(R_{\rm p}+R_{\rm t})}$.  A bulk density of
3~g~cm$^{-3}$ was used in the two studies considered here, so we adopted
the same value to calculate $R_{\rm p}$ and $R_{\rm t}$ in this work.

For grazing collisions with impact velocities exceeding $V_{\rm hr}$,
the outcome transitions to the hit-and-run regime. In a hit-and-run
event, the target retains approximately its original mass (although
there may be some material exchange between the two bodies) but the
escaping projectile may be disrupted or intact. Projectile disruption
is calculated for the reverse impact situation: the interacting mass
of the target onto the total projectile mass. The interacting mass of
the target is approximated by the overlapping cross sectional area of
the projectile times the chord length through the target. The rest of
the target mass is assumed to escape completely and is neglected in
the reverse calculation.  The same procedure is used to determine the
disruption criteria for the reverse problem as described above; the
full derivation of the reverse impact is presented in LS12. When the
projectile mass is less than about 10\% of the target, the projectile
experiences disruption during most hit-and-run events.

For grazing collisions with impact velocities approaching $V^{\prime
  *}$, the collision outcome transitions from hit-and-run to erosion
of the target. For such grazing collisions in the disruption regime,
the outcome is defined by the catastrophic disruption criteria
$Q^{\prime *}_{RD}$ and Equation \ref{eqn:univlaw} or
\ref{eqn:mlrsuper} only for $M_{\rm lr}<M_{\rm t}$.

In the disruption regime, recent hydrocode simulations of collisions
between differentiated bodies showed that the core is preferentially
incorporated into the largest remnant
\citep{Marcus:2009,Marcus:2010}. Consider two idealized models:
\begin{enumerate}
\item{Model 1 -- Cores always merge: Given the original
    $M_{\rm core,t}$ and $M_{\rm core,p}$, the post-impact core is
    $M_{\rm core,lr}={\rm min}(M_{\rm lr},M_{\rm core,t}+M_{\rm core,p})$.}
\item{Model 2 -- Cores only merge on accretion: When
    $M_{\rm lr}>M_{\rm t}$, $M_{\rm core,lr}=M_{\rm core,t}+{\rm
      min}(M_{\rm core,p},M_{\rm lr}-M_{\rm t}$). When $M_{\rm
      lr}<M_{\rm t}$,
    assume that none of the projectile accretes and the mantle is
    stripped first. Then, $M_{\rm core,lr}={\rm min}(M_{\rm
      core,t},M_{\rm lr})$.}
\end{enumerate}
\citet{Marcus:2010} found that the core mass fraction in the largest
remnant in the hydrocode simulations fell between these two idealized
models. Hence, in this work, we averaged the predicted core fractions
from the two models to calculate the change in core mass fraction,
$f_{\rm core}$. We assumed that the initial $f_{\rm core}=1/3$ to be
comparable to Earth.

Collision outcome maps are shown in Figure \ref{fig:collmap} for a
mass ratio of 1:10, which is typical for the late giant impacts onto
Earth-mass planets, and a mass ratio of 1:40, representing the initial
mass ratio of planetesimals colliding with embryos in the $N$-body
simulations discussed below.  The impact velocity is normalized to the
mutual surface escape velocity, and the impact angle axis is spaced by
equal probability according to $\sin \theta \cos \theta d \theta$
\citep{Shoemaker:1962}. Collision outcomes are divided into 6 groups
(Figure~\ref{fig:collmap}):
\begin{enumerate}
\item{Perfect merging when $V_i<V^{\prime}_{\rm esc}$ (dark blue
    region in Figure~\ref{fig:collmap}),}
\item{Graze-and-merge when $b>b_{\rm crit}$ and
    $V^{\prime}_{\rm esc}<V_i<V_{\rm hr}$
    (blue-green region),}
\item{Hit-and-run when $b>b_{\rm crit}$, $V_i>V_{\rm hr}$ and
    $M_{\rm lr}=M_{\rm t}$ (green region),}
\item{Partial accretion when $b<b_{\rm crit}$ and $M_{\rm lr}>M_{\rm t}$
    (medium blue region),}
\item{Partial erosion when $0.1M_{\rm tot}<M_{\rm lr}<M_{\rm t}$ (e.g.,
    thick black line denotes $0.5M_{\rm tot}$),}
\item{Super-catastrophic when $M_{\rm lr}<0.1M_{\rm tot}$.}
\end{enumerate}
Note that collisions with impact parameters near $b_{\rm crit}$ are
difficult to predict (red vertical lines in Figure~\ref{fig:collmap}),
and outcomes near $b_{\rm crit}$ display a mix of features from both
the accretion-to-erosion regimes and the merging-to-hit-and-run
regimes.

In the collision outcome maps, contours of constant final mass are
shown by deriving the corresponding impact velocity as a function of
impact angle, $V_{i,\rm lr=const}$. Using the angle-dependent
$Q^{\prime *}_{RD}$ for the desired mass ratio, the impact energy
contour is
\begin{equation}
Q_{R,\rm lr=const} =  Q^{\prime *}_{RD} \left [ -2 \left( \frac{M_{\rm
        lr}}{M_{\rm tot}}={\rm const} \right) + 2 \right ], 
\end{equation}
using the universal law (Equation \ref{eqn:univlaw}) for $0.1<M_{\rm
  lr}/M_{\rm tot}<1$. For super-catastrophic events, the 
impact energy contour is
\begin{equation}
  Q_{R,\rm lr=const} =  Q^{\prime *}_{RD} \left (
    \frac{1.8^{\eta}}{0.1} \left( \frac {M_{\rm lr}}{M_{\rm tot}}
      ={\rm const} \right) \right )^{1/\eta}.
\end{equation}
from Equation \ref{eqn:mlrsuper}.  Using Equation \ref{eqn:qr}, the
corresponding impact velocity as a function of impact angle is given
by
\begin{equation}
V_{i,\rm lr=const} = \sqrt{ 2 Q_{R, \rm lr=const} M_{\rm tot} / \mu }.
\end{equation}

The onset of collisional erosion is determined by the value of the
material parameter $c^*$. With $c^*=1.9$ and the initial mass ratios
in typical $N$-body simulations, erosion begins with impact velocities
of about 2 to 3 times the mutual escape velocity for head-on
collisions and increases with impact parameter
(Figure~\ref{fig:collmap}).  For collisions between similarly sized
bodies, the transition between accretion and catastrophic disruption
occurs over a very small range in impact velocity.  For example, the
critical impact velocities required to begin eroding and to
catastrophically disrupt an embryo by a body half its mass are only
$2.0V_{\rm esc}$ and $2.5V_{\rm esc}$ at $b=0$, respectively. As the
mass ratio becomes more extreme, the impact velocities required to
reach catastrophic disruption increase significantly.  In Figure
\ref{fig:collmap}A, erosion begins at $2.5V_{\rm esc}$ and disruption
at $5.9V_{\rm esc}$ for a mass ratio of 1:10.  For the 1:40 mass ratio
shown in Figure~\ref{fig:collmap}B, the disruption of an embryo by a
planetesimal would require an impact velocity of at least $18V_{\rm
  esc}$.

The contours of constant largest remnant have a different shape for
the reverse impact in the hit-and-run regime compared to the forward
scenario. For impact parameters near $b_{\rm crit}$ and $M_{\rm
  p}/M_{\rm t} \le 0.1$, the interacting mass of the target is
approximately equal to the projectile mass. As the impact angle
increases, the reverse projectile-to-target mass ratio
decreases. Hence, the total interacting mass in the reverse impact
decreases with increasing impact angle.  The velocity contours
correspond to constant remnant mass divided by total interacting
mass. The changing total mass and mass ratio leads to the intersection
of the catastrophic disruption contour (blue dot-dashed lines in
Figure~\ref{fig:collmap}) and onset of projectile erosion contour
(blue dotted lines) near $b_{\rm crit}$ and divergence at larger
impact angles. Futhermore, there is a minimum in the projectile
erosion contour at an optimum interacting mass fraction from the
target.

\section{Analysis of $N$-body Planet Formation Simulations}

We examined the collisions in two recent $N$-body studies of the late
stages of terrestrial planet formation.  \citet{Obrien:2006} performed
8 simulations of terrestrial planet formation beginning with 25
Mars-mass embryos and an equal mass of material in a population of
1000 0.00233-$M_{\rm Earth}$ planetesimals in a zone from 0.3 to 4
AU. \citet{Raymond:2009} performed 40 simulations beginning with 85-90
planetary embryos (0.005 to 0.1 $M_{\rm Earth}$) and 1000 or 2000
0.0025-$M_{\rm Earth}$ planetesimals between 0.5 and 4.5 AU. 

The beginning stage of each study approximates the end of oligarchic
growth and the onset of stochastic growth. When the total mass in
planetary embryos is equal to the total mass in smaller planetesimals,
viscous stirring by the embryos cannot be damped by dynamical friction
from the planetesimals, which marks the end of oligarchic growth
\citep{Kokubo:1998,Goldreich:2004}. It was assumed that the nebular gas
had dissipated and Jupiter and Saturn were fully formed at the start of
the simulations. The orbital configurations of Jupiter and Saturn were
varied over a plausible range to consider their influence on the final
state of planets in the terrestrial zone. Not all cases result in
terrestrial planets in agreement with our Solar System; moreover, the
dynamical history of the giant planets in our Solar System is still an
active area of research with new ideas that have not yet been
incorporated into detailed accretion studies of the terrestrial
planets \citep[e.g.,][]{Walsh:2011,Morbidelli:2010}. Hence, the set of
simulations are representative of the general dynamics of terrestrial
planet growth in the presence of two outer gas giant planets.

The study by \citet{Obrien:2006} used the SyMBA $N$-body code
\citep{Duncan:1998}, and \citep{Raymond:2009} used the Mercury
$N$-body code \citep{chambers:1999}. Both codes are similar symplectic
integrators that calculate close encounters between planetary
bodies. For every collision, the two bodies were merged and linear
momentum was conserved. In these simulations, the embryos
gravitationally interacted with all other bodies. The gravitational
influence of planetesimals upon embryos was calculated; however,
planetesimals did not influence other planetesimals and could not
collide with each other. In this work, we present a retrospective analysis
of the simulations to assess the range of true collision outcomes
during the end stage of terrestrial planet formation.

\subsection{Collision outcomes}

\begin{table*}[t]
  \caption{\begin{flushleft}Predicted collision outcome statistics
      from recent $N$-body simulations of the end stage of terrestrial
      planet formation using the new collision physics model with
      material parameters $c^*=1.9$ and $\bar \mu=0.36$. The $N$-body
      simulations began with a population of planetary embryos and
      non-interacting planetesimals and all collisions resulted in
      perfect merging. Giant impacts are collisions between embryos. \end{flushleft}}
  \footnotesize
  \begin{tabular}{l || rr | rr || rr | rr || rr | rr } \hline \hline
    & \multicolumn{4}{c||}{Group 1}                                      & \multicolumn{4}{c||}{Group 2}                                     & \multicolumn{4}{c}{Group 3} \\
    & \multicolumn{4}{c||}{O'Brien et al. 2006}                          & \multicolumn{4}{c||}{Raymond et al. 2009}                         & \multicolumn{4}{c}{Raymond et al. 2009} \\
    & \multicolumn{4}{c||}{15 Large Planets from 8 Sims.}              & \multicolumn{4}{c||}{52 Large Planets from 40 Sims.}            & \multicolumn{4}{c}{161 Total Planets from 40 Sims.} \\
    & \multicolumn{4}{c||}{$0.74-1.58 M_{\rm Earth}$}                       & \multicolumn{4}{c||}{$0.70-1.45 M_{\rm Earth}$}                      & \multicolumn{4}{c}{$0.05-1.45 M_{\rm Earth}$} \\
    & \multicolumn{2}{c|}{Planetesimal}  & \multicolumn{2}{c||}{Giant}    & \multicolumn{2}{c|}{Planetesimal}  & \multicolumn{2}{c||}{Giant}  & \multicolumn{2}{c|}{All Giant} & \multicolumn{2}{c}{Last Giant} \\ 
    Collision outcome  & $N=1140$ & \% & $N=67$ & \% & $N=3142$ & \% & $N=544$ & \% & $N=1165$  & \% & $N=161$  & \%    \\ \hline 
    Super-catastrophic & 0      & 0    & 1    & 1    & 0    & 0      & 0    & 0     & 4    & $<1$    & 3  & 2  \\
    Partial erosion    & 0      & 0    & 1    & 1    & 61   & 2      & 3    & $<1$  & 11   & $<1$    & 5  & 3 \\
    Partial accretion  & 828    & 73   & 18   & 27   & 2180 & 69     & 213  & 39    & 421  & 36      & 62 & 39 \\
    Perfect merging    & 0      & 0    & 0    & 0    & 18   & $<1$   & 4    & $<1$  & 7    & $<1$    & 0  & 0 \\
    Graze-and-merge    & 43     & 4    & 26   & 39   &  85  & 3      & 173  & 32    & 394  & 34      & 31 & 19 \\
    Hit-and-run (H\&R)   & 269    & 24   & 21   & 31   & 798  & 25     & 151  & 28    & 328  & 28      & 60 & 37 \\ \hline \hline
    Special cases      &        &      &      &      &      &        &     &        &      &         &     & \\
  H\&R with proj. erosion         & 253  & 22 & 2 & 3  & 778 & 25    & 75   & 14    & 138  & 12      & 35 & 22  \\
  $5-10$\% increase in $f_{\rm core}$ &  0 & 0  & 7 & 10 & 0   & 0     & 132  & 24    & 128  & 11      & 14 & 9   \\ 
  $>10$\% increase in $f_{\rm core}$  &  0 & 0  & 4 & 6  & 2  & $<1$   & 90   & 17    &  75  & 6       & 19 & 12 \\ \hline \hline
 & & \\
\end{tabular}
\label{tab:outcomes}
\end{table*}

We divided the collision outcomes from the $N$-body studies into three
groups, described in Table~\ref{tab:outcomes}. Group 1 consists of all
collisions leading to the growth of 15 planets with final masses of
0.74 to 1.58 $M_{\rm Earth}$ from 8 different simulations in
\citet{Obrien:2006}. These 15 planets were also studied by
\citet{Nimmo:2010} to investigate the evolution of the
hafnium-tungsten isotopic system during planet growth. Group 2
includes all collisions leading to the growth of 52 planets with final
masses between 0.70 and 1.45 $M_{\rm Earth}$ from 40 simulations by
\citet{Raymond:2009}. Group 3 considers only the giant impacts onto all 
161 planets from the 40 simulations in \citet{Raymond:2009}. Giant
impacts are defined as collisions between planetary embryos. A final
planet experiences at least one giant collision; hence, group 3
excludes surviving embryos that only accrete planetesimals (26 embryos
survived without a giant impact in 40 simulations).

In each case, the simulation collision parameters were used to
calculate the outcome based on our analytic model. The outcome depends
on the mass ratio of the two bodies, the impact angle, the impact
velocity normalized to the mutual escape velocity, and the
catastrophic disruption criteria.  Here, we used values of $c^*=1.9$
and $\bar \mu=0.36$ to calculate the catastrophic disruption criteria,
which are appropriate for strengthless planets as we assumed that the
planets are hot and possibly partially molten during the late stage of
planet formation.

\subsubsection{Group 1}

\begin{figure*}[t]
\begin{center}
\includegraphics[width=35pc]{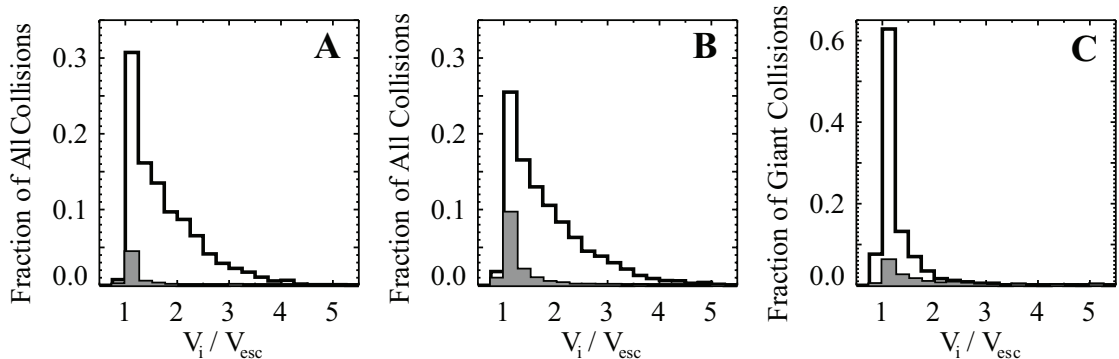}
\end{center}
\caption{Distribution of impact velocities normalized to mutual
  surface escape velocity for (A) all collisions (black line) and all
  giant impacts (filled grey) in group 1; (B) all collisions (black
  line) and all giant impacts (filled grey) in group 2; (C) all giant
  impacts (black line) and last giant impact (filled grey) in group
  3 $N$-body simulations.\\}
\label{fig:vimpact}
\end{figure*}

There were a total of 1207 collisions by planetesimals and embryos
during the growth of the 15 planets in group 1. Of these, 1140
collisions were by planetesimals.  The analytic model predicts that
the majority of planetesimal collisions will lead to partial accretion
(73\%). The dominance of accretionary events is foremost a reflection
of the impact velocity distribution, which is peaked between 1 and
$2V_{\rm esc}$ (Figure~\ref{fig:vimpact}A). Note that only half of the
planetesimal is accreted onto an embryo when $V_i=2V_{\rm esc}$ (the
critical velocity for accretion of half the projectile rises slightly
with increasing impact angle, see Figure~\ref{fig:collmap}B). The
distribution of accretion efficiencies \citep[$(M_{\rm lr}-M_{\rm
  t})/M_{\rm p}$) from ][]{asphaug:2009} is shown in
Figure~\ref{fig:acceff}A. The most common outcome of a planetesimal
encounter was accretion of $>90$\% of the mass, although there were a
significant number of events where less material was accreted. For
collisions between embryos, a smaller fraction of the projectile is
accreted (70 to 90\% for the 18 partial accretion events).

Beginning with Mars-sized embryos, there were no cases of erosion of
the growing planet from planetesimal encounters. Notably, a
substantial fraction of planetesimal encounters were hit-and-run
events. For the initial 1:43 mass ratio between the planetesimals and
embryos, the critical impact parameter is about 0.78
(51$^{\circ}$). Given the probability distribution of impact angles,
about 40\% of outcomes for 1:43 mass ratio collisions are in the
grazing regime, and the probability decreases as the embryos grow and
the mass difference increases. Of all the impacts by planetesimals
during the growth of Earth-mass planets, about 4\% were
graze-and-merge and 24\% were hit-and-run. Erosion of the planetesimal
occurred in nearly all of the hit-and-run collisions (253 out of 269),
and catastrophic disruption of the planetesimal occured in about 85\%
of these events (Figure~\ref{fig:collmap}B).

Next, we consider only the giant impacts in group 1. Giant collision
outcomes are approximately evenly split between partial accretion,
graze-and-merge, and hit-and-run. Only a few of the embryo
projectiles in hit-and-run events are eroded; in these collisions, the
target and projectile have comparable masses ($\gamma \ge 0.1$) and
neither body is disrupted.

The largest impact velocities ($>6V_{\rm esc}$), although rare, are
high enough for embryos to catastrophically disrupt each other.  There
are notable examples of partial erosion (planet CJS1.4) and
super-catastrophic (planet EJS1.4) outcomes. The last giant impact (at
222 Myr) onto a $1.05 M_{\rm Earth}$ planet by a $0.1 M_{\rm Earth}$
embryo at $3.2V_{\rm esc}$ and $35^{\circ}$ resulted in erosion of
about 1\% of the target mass (Figure~\ref{fig:collmap}A). The second
giant impact (at 9.7 Myr) onto a $0.21 M_{\rm Earth}$ body by a $0.147
M_{\rm Earth}$ embryo at $6.2V_{\rm esc}$ and $26^{\circ}$
super-catastrophically disrupted the target leaving a largest remnant
of only $0.008 M_{\rm Earth}$.

\subsubsection{Group 2}

\begin{figure*}[t]
\begin{center}
\includegraphics[width=35pc]{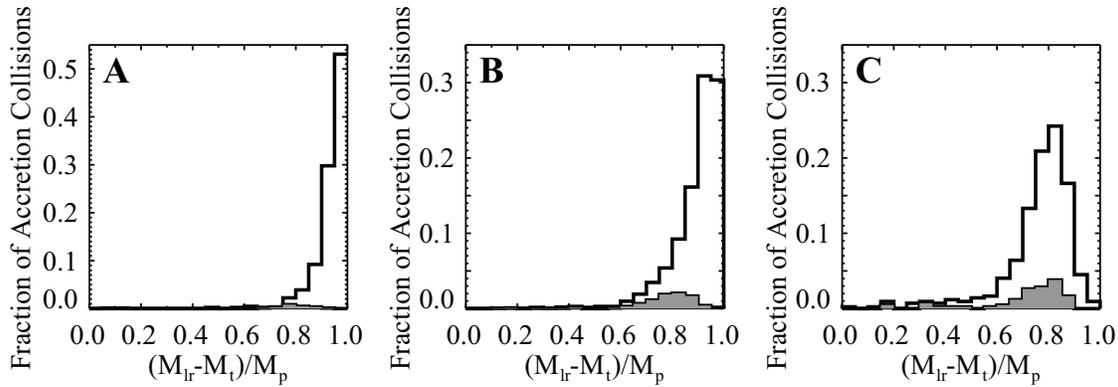}
\end{center}
\caption{Distribution of accretion efficiency ($(M_{\rm lr}-M_{\rm t})/M_{\rm p}$)
  for collisions that result in partial accretion: (A) all collisions
  (black line) and giant impacts (filled grey) in group 1; (B) all
  collisions (black line) and all giant impacts (filled grey) in group
  2; (C) all giant impacts (black line) and last giant impact (filled
  grey) in group 3 $N$-body simulations.}
\label{fig:acceff}
\end{figure*}

There were a total of 3686 collisions by planetesimals and embryos
during the growth of the 52 planets in group 2. Of these, 3142
collisions were by planetesimals. As in group 1, the analytic model
predicts that about 70\% of planetesimal collisions will lead to
partial accretion. The simulations by \citet{Raymond:2009} considered
a wider range of dynamical configurations for Jupiter and Saturn,
which produced a slightly wider distribution of impact velocities
(Figure~\ref{fig:vimpact}B) compared to group 1. In addition, the
initial masses of the embryos was smaller. As a result, a couple of
percent of planetesimals impacts led to erosion of the growing planet.
For collisions in the partial accretion regime, the mean accretion
efficiency is slightly lower in group 2 and the tail of low efficiency
events is more pronounced than in group 1 (Figure~\ref{fig:acceff}B).

Overall, the probabilities of different collision outcomes for
planetesimal impacts are similar in groups 1 and 2 because of the
similar mass ratios and impact velocity distributions
(Table~\ref{tab:outcomes}). Again, most of the planetesimal
hit-and-run events result in catastrophic disruption of the
projectile.

\begin{figure*}
\begin{center}
\includegraphics[width=35pc]{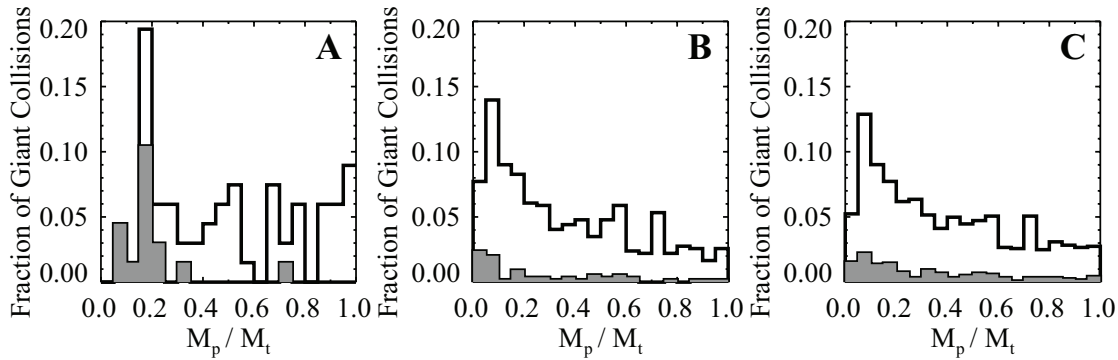}
\end{center}
\caption{Distribution of projectile-to-target mass ratio for all giant
  impacts (black lines) and last giant impact (filled grey) for (A)
  group 1; (B) group 2; (C) group 3 $N$-body simulations. }
\label{fig:gamma}
\end{figure*}

Compared to the giant impacts in group 1, group 2 giant impacts have
more partial accretion events and significantly more embryos are
eroded in hit-and-run events. The difference is primarily a result of
the fact that the initial embryos were smaller in the simulations by
\citet{Raymond:2009}. The larger mass ratio between the embryos and
growing planet leads to more cases of fragmentation of the smaller
body and fewer grazing impacts. 22\% of giant impacts in group 2 have
$\gamma < 0.1$, but only 4\% of group 1 giant impacts have such a
large mass contrast (Figure~\ref{fig:gamma}).

Of the 151 hit-and-run giant impacts in group 2, 75 projectiles were
eroded (50\%), and 29 projectiles suffered catastrophic disruption
level fragmentation (19\%). The three erosive giant impacts in this
group removed 16, 6, and 1\% of the material from target bodies with
initial masses of 0.08, 0.50 and $0.72 M_{\rm Earth}$,
respectively. The erosive events all occurred in the eccentric Jupiter
and Saturn (EJS) group of simulations with impact velocities between
2.1 and $3.2 V_{\rm esc}$.

Because partial accretion is the most common outcome of non-grazing
collisions, a significant fraction of giant impacts result in
potentially observable changes in the bulk composition of a planet
(Table~\ref{tab:outcomes}). For collisions between differentiated
bodies, the core-to-mantle mass ratio changes during both partial
accretion and erosion events. For the large final planets in group 2,
a 5\% or greater increase in the mass fraction of the core, $f_{\rm
  core}$, occured in about 41\% of all giant impacts and in about 20\%
of last giant impacts.

\subsubsection{Group 3}

Considering all 1165 giant impacts onto 161 planets in 40 planet
formation simulations, the outcomes are approximately evenly split
between partial accretion, graze-and-merge, and hit-and-run. Erosive
events, including super-catastrophic disruption, occur about 1\% of
the time.  In the hit-and-run events, about half of the projectiles
are eroded and about 20\% are catastrophically disrupted.

\begin{figure*}[t]
\begin{center}
\includegraphics[width=35pc]{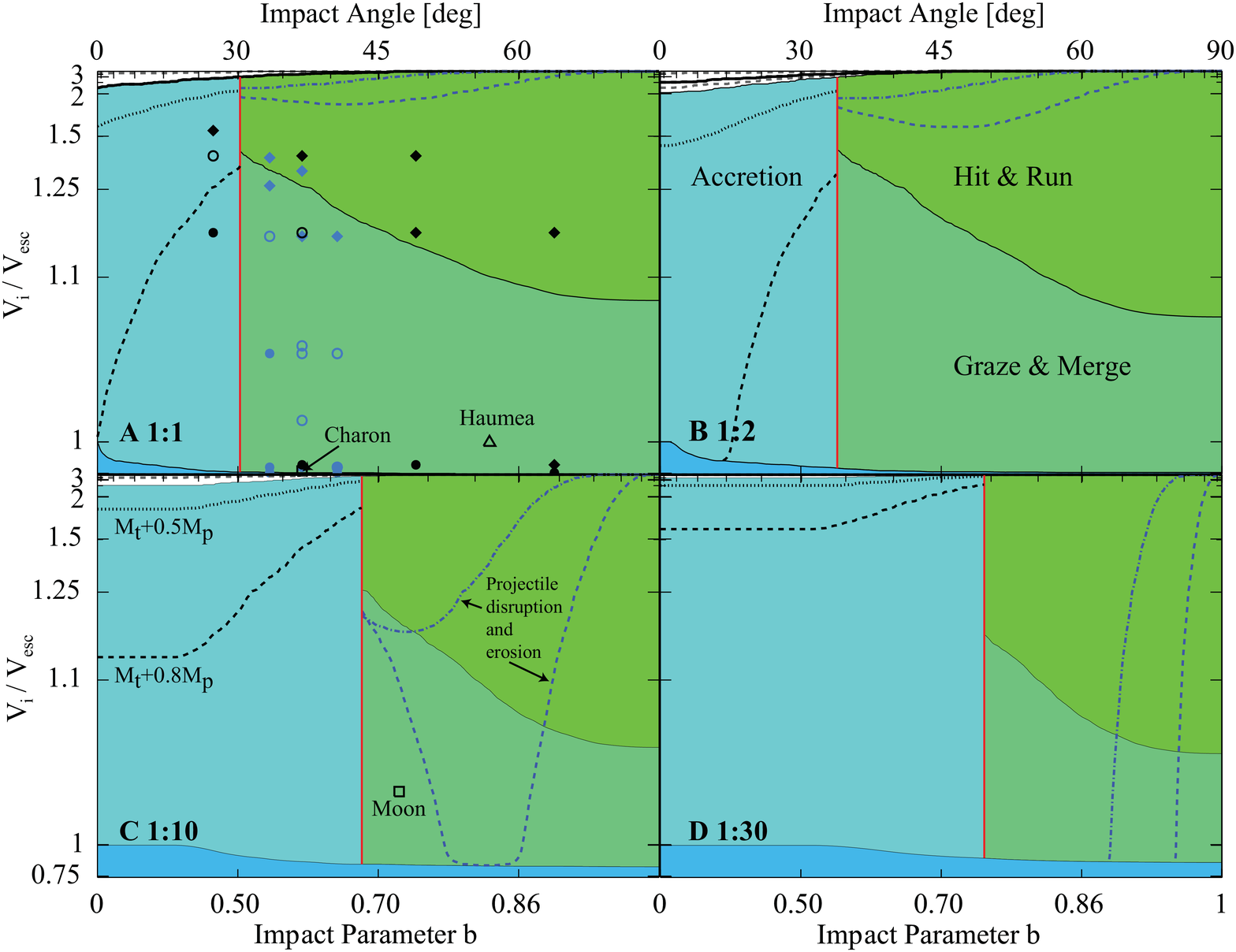}
\end{center}
\caption{Collision outcome maps with area scaled by outcome
  probability. Collision maps for planetary embryos, using material
  parameters $c^*=1.9$, $\bar \mu=0.36$, for projectile-to-target mass
  ratios of (A) $M_{\rm p}:M_{\rm t}=1:1$, (B) 1:2, (C) 1:10, and (D)
  1:30. Vertical axis is scaled by velocity distribution of giant
  impacts in group 3 $N$-body simulations (Figure~\ref{fig:vimpact}).
  Colors are the same as in Figure~\ref{fig:collmap}.  Dotted black
  curve -- 50\% of projectile accreted; dashed black curve -- 80\% of
  projectile accreted; dot-dashed blue curve -- catastrophic
  disruption of projectile in a hit-and-run; dashed blue curve --
  onset of erosion of projectile in a hit-and-run. Example results
  from numerical simulations: filled circles are merging, open circles
  and graze-and-merge, diamons are hit-and-run. Black -- planetesimal
  collisions from \citet{Leinhardt:2000}; blue -- Haumea formation
  simulations and favored Haumea impact scenario (triangle) from
  \citet{Leinhardt:2010}; arrow -- disk-origin scenario for Charon
  from \citet{Canup:2005}; square -- typical moon formation scenario
  from \citet{Canup:2004}. \\}
\label{fig:collmapequal}
\end{figure*}

The equal likelihood of partial accretion, graze-and-merge, and
hit-and-run is a result of the range of mass ratios and impact
velocities for giant impacts. Given the impact velocity distribution
in group 3, the velocity axis of a collision outcome map may be scaled by probability. In
Figure \ref{fig:collmapequal}, both axes are scaled by probability;
hence, the area of each collision outcome (denoted by colors) is
directly proportional to their probability. 

The 4 panels span the range of giant impact mass ratios in group 3
(99\% of events have $\gamma>0.03$ but a few embryo-embryo collisions
have mass ratios as extreme as 1:55). Note that, over the course of
the entire simulation, the mass ratio of a giant impact is about
equally likely to fall anywhere between 0.03 and 1
(Figure~\ref{fig:gamma}C). Even when considering just the largest
final planets (group 2), giant impacts may be any mass ratio (e.g., a
collision between two $0.4 M_{\rm Earth}$ bodies). However, the last
giant impact onto target bodies greater than about $0.8 M_{\rm Earth}$
are dominated by mass ratios less than 0.1
(Figure~\ref{fig:group3vels}F).

The collision maps fully scaled by probability emphasize the
importance of the graze-and-merge regime even though it is a narrow
regime in absolute impact velocity. The scaled figures also emphasize
that partial accretion of the projectile is the most common outcome
for non-grazing collisions; recall that the accretion efficiency peaks
at about 80\% for all giant impacts (Figure~\ref{fig:acceff}C). As
shown in Figure \ref{fig:collmapequal}, hit-and-run events occur about
1/3 of the time and the projectile is eroded when the projectile mass
is less than about 10\% of the target.

Note that the boundary between graze-and-merge and the adjacent
partial accretion and hit-and-run regimes derived by
\citet{Kokubo:2010} is generally in good agreement with other
simulations with $\gamma=1$. However, the simulations of rubble pile
collisions with $\gamma=0.5$ using the pkdgrav code
\citep{Leinhardt:2010} found a much narrower graze-and-merge regime
compared to the boundary derived from SPH simulations of fluid
bodies. More work is needed to understand the boundaries of the
graze-and-merge regime and its dependence on material properties.

As shown in Table~\ref{tab:outcomes}, the last giant impact was more
likely to be erosive (5\%) or to be a hit-and-run (37\%) compared to
all giant impacts. The dynamical stirring by the last remaining
planets leads to higher impact velocities near the end of planet
formation compared to the total time average. In
Figure~\ref{fig:vimpact}C, note that the distribution of impact
velocites is more weighted to values $>1.25 V_{\rm esc}$ for the last
impact (grey filled histogram) compared to all giant impacts (black
histogram). For the same reason, more of the projectiles in the last
giant hit-and-run event are eroded compared to all giant impacts (35
out of 60 events).

\subsection{Collisions between planetesimals}

\begin{figure*}[t]
\begin{center}
\includegraphics[width=40pc]{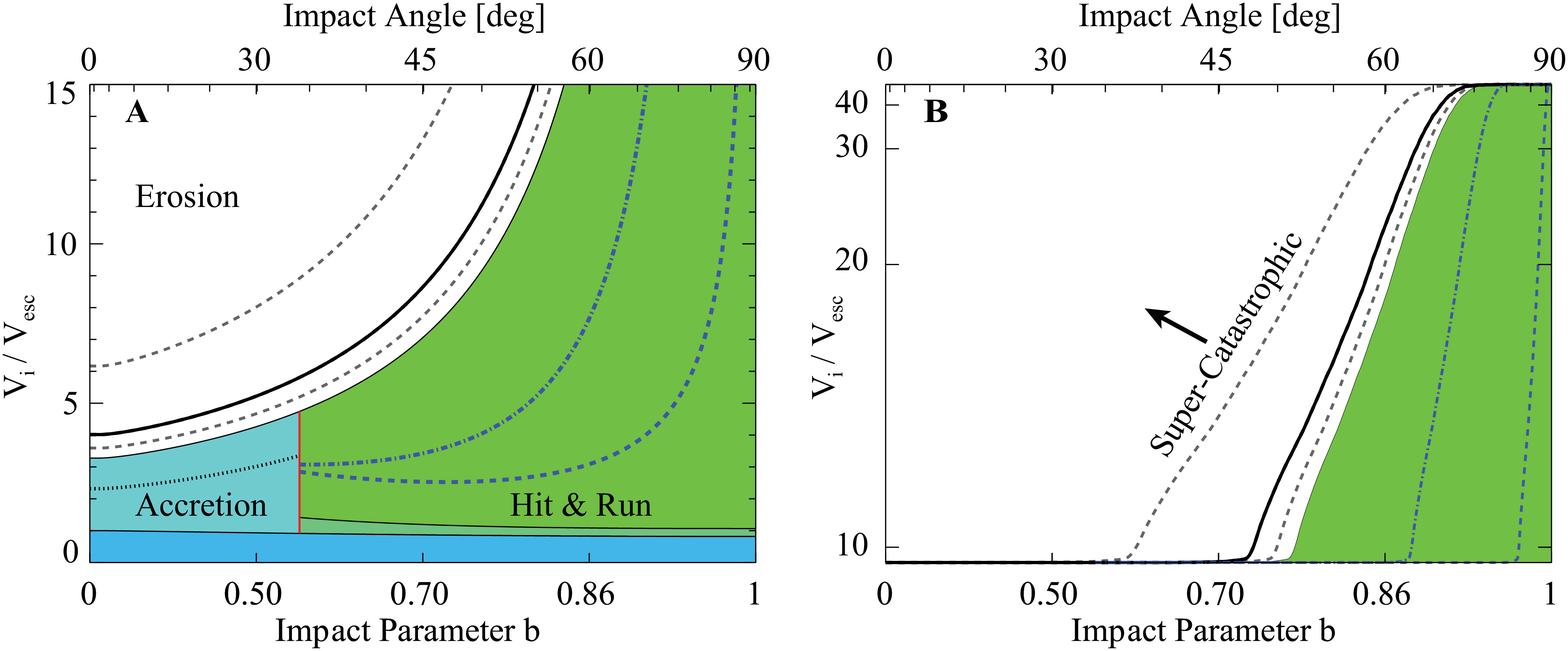}
\end{center}
\caption{Collision outcome maps for collisions between planetesimals with
  projectile-to-target mass ratio $M_{\rm p}:M_{\rm t}=1:2$ and
  material parameters of $c^*=5$ and $\bar \mu=0.37$, which are
  appropriate for a wide range of solid compositions. (A) Map with
  linear impact velocity normalized to planetesimal escape velocity
  and impact angle scaled by probability, and (B) map with area scaled
  to outcome probability for typical collision velocities stirred up
  by the presence of growing planets. Colors are the same as in
  Figure~\ref{fig:collmap}.  Thick black curve -- critical velocity
  for catastrophic disruption ($M_{\rm lr}=0.5M_{\rm tot}$); dashed
  grey curves -- 10\% and 90\% of target mass in largest remnant;
  dotted black curve -- 50\% of projectile accreted; dot-dashed blue
  curve -- catastrophic disruption of projectile in a hit-and-run;
  dashed blue curve -- onset of erosion of projectile in a
  hit-and-run. \\}
\label{fig:collmapp}
\end{figure*}

Note that the collisions considered in groups 1--3 are only those that
contributed to the formation of the final planets. At the beginning of
the simulation, there should have been many collisions between
planetesimals, but they were not modeled. The collision outcome map
for impacts between planetesimals with a mass ratio of 1:2 is shown in
Figure~\ref{fig:collmapp}. Typical planetesimal-planetesimal impact
velocities will be similar to their collision velocities onto the
embryos (about the escape velocity from the embryo,
Figure~\ref{fig:vimpact}). The surface escape velocity from an
Earth-mass body is about 10 times the escape velocity from a
$0.0025M_{\rm Earth}$ planetesimal. Erosion during collisions between
comparable mass planetesimals begins at about $3.6V_{\rm esc}$ for
$c^*=5$, a value appropriate for a variety of solid compositions.
Hence, the most common planetesimal-planetesimal collision outcome is
super-catastrophic destruction of the planetesimals.  The
probability-scaled collision map (Figure \ref{fig:collmapp}B)
emphasizes the extremely destructive nature of
planetesimal-planetesimal collisions in the presence of growing
planets.

As shown in Figure \ref{fig:collmapp}B, collisions more oblique than
about $60^{\circ}$ are hit-and-run.  If planetesimal-planetesimal
collisions had been modeled in the $N$-body simulations, the
probability of a hit-and-run would have been artifically high at the
beginning of the calculation because of the assumed starting
distribution of equal-mass planetesimals. In fact, planetesimals in a
size distribution defined by a collisional cascade would have a
smaller fraction of mutual hit-and-run events. The ultimate fate of
the smallest fragments in the collisional cascade is determined by the
competition between accretion onto the growing planets and removal
from the planet's feeding zone (e.g., via Poynting-Robertson drag).

In the $N$-body simulations, the planetesimals all had the same mass
because of computational limitations that restrict the total number
particles. A tractable number of particles (few thousand) was
insufficient to resolve a size distribution of planetesimals. However,
the fraction of planetesimal collisions onto embryos that lead to
partial accretion vs.\ hit-and-run depends on the mass ratio. If more
of the mass in planetesimals were in smaller bodies, then accretionary
collisions would be more common than found in the $N$-body simulations
considered here.

\section{A Monte Carlo Planet Growth Model}

The restrospective analysis of $N$-body simulations presented above
provides a limited view into the role of more realistic collision
physics on planet growth because it cannot assess the cumulative
effects of different collision outcomes. Studying suites of Monte
Carlo simulations of the growth of a single planet via giant
collisions allows for an intermediate examination of the role of
collisions that is more tractable than many new full $N$-body
simulations. The purpose of this Monte Carlo simulation is to
investigate the effects of the collision physics model and is not
meant to replace full $N$-body planet formation models or more
sophisticated statistical population synthesis models
\citep[e.g.,][]{Alibert:2011,Mordasini:2009,Mordasini:2009b,Ida:2010}. 

\subsection{Method}
Our planet growth model assumed that the overall dynamics of giant
impacts is similar to the $N$-body simulations by
\citet{Raymond:2009}. Because planetary embryos become dynamically
isolated when their masses are much smaller than an Earth-mass planet
(roughly one tenth the mass), most of the mass in large terrestrial
planets, here defined as a final mass $\ge 0.7M_{\rm Earth}$, is
accreted via giant impacts \citep{Kokubo:1998}. Thus, the Monte Carlo
simulation uses the distribution of impact parameters from the group 3
$N$-body simulations to model planet growth (summarized in Figures
\ref{fig:group3} and \ref{fig:group3vels}). In the group 3 planets,
the fraction of mass accreted via planetesimal collisions varied from
0 to 48\%. The distribution of mass accreted from planetesimals is
sensitive to the final mass of the planet; the large planets accreted
between 2 and 27\% of their final mass from planetesimals in the
stochastic stage of planet formation modeled in the $N$-body
simulations (Figure \ref{fig:group3}).

For each planet, the following procedure was applied:
\begin{enumerate}
\item{Randomly choose a value from the probability distribution of
    initial masses (Figure~\ref{fig:group3}). The minimum initial
    embryo mass was $0.01 M_{\rm Earth}$.}
\item{Randomly choose a value from the probability 
    distribution of number of giant impacts (Figure~\ref{fig:group3}).}
\item{For each giant impact, randomly choose the projectile mass,
    impact velocity, and impact angle. Projectile mass and impact
    velocity distributions are dependent on target mass
    (Figure~\ref{fig:group3vels}).}
\item{Three sets of simulations were run: (i) perfect merging for all
    collisions for comparison to the $N$-body results; (ii) collision
    physics assuming no re-impact by the projectile in hit-and-run
    events (group A); and (iii) collision physics assuming re-impact
    by the projectile in hit-and-run events (group B).}
\item{Add a randomly chosen value from the probability distribution of
    mass contributions from planetesimals. The planetesimal
    contribution distribution is also dependent on target mass
    (Figure~\ref{fig:group3}).}
\end{enumerate}

For each simulation set, we modeled the growth of 200 planets. When
using the collision physics model, we tracked the core mass fraction
assuming that all embryos have an initial value of 1/3 to be
comparable to the bulk Earth.  The core mass fraction of planetesimals
was assumed to be the same as the initial embryos. We recorded the
mass of debris produced by each giant impact. For the purpose of
comparing perfect merging and the collision model, we assumed that all
debris was lost and not reaccreted later. This end member assumption
represents the maximum effect of fragmentation on inhibiting planet
growth in the giant impact stage.

The distributions of impact velocities and projectile-to-target mass
ratios are significantly dependent on the target mass (see
Figure~\ref{fig:group3vels}). The population of initial embryos
collides with each other to grow the first mid-size planets.  The most
common collision mass ratio depends on the initial size distribution
of embryos assumed in a particular $N$-body simulation.  During this
early period of growth in the group 3 simulations, the mass ratios of
the two bodies was about equally distributed between $0<\gamma<1$
(Figure~\ref{fig:group3vels}D).  At the same time, the impact
velocities are strongly peaked just above $1V_{\rm esc}$
(Figure~\ref{fig:group3vels}A) and few collisions lead to erosion
(Figure~\ref{fig:group3vels}G). As the planets continue to grow,
impacts onto the larger bodies are dominated by smaller embryos, and
the distribution of mass ratios is strongly peaked with $\gamma<0.2$
for Earth-mass targets (Figure~\ref{fig:group3vels}F).  The largest
planets are strong perturbers on the remaining small embryos and the
distribution of impact velocities becomes wider compared to earlier
planet growth, spanning $1-3V_{\rm esc}$
(Figure~\ref{fig:group3vels}C).

If these mass-dependent effects were not included and the mean
distributions were utilized instead, then the Monte Carlo model would
predict too many very large planets ($>1.5 M_{\rm Earth}$). In other
words, the very end of terrestrial planet growth has distinct
dynamical differences from earlier parts of the stochastic giant
impact phase. To account for the variations with time, the values for
impact velocity and mass ratio were chosen from the subset of group 3
collisions within $\pm0.1 M_{\rm Earth}$; when the target had a mass
greater than one Earth mass, the distribution was based on the group 3
data with $M_{\rm t}>0.9M_{\rm Earth}$.

The projectile mass may be eroded during hit-and-run
events. Therefore, in the group B set, we calculated the mass of the
largest projectile remnant using the reverse collision scenario. The
returning projectile had the mass of the largest remnant and the rest
of the projectile debris was neglected. The re-impact had a randomly
chosen impact angle and an impact velocity given by the greater of
$V_i \sin \theta$ or $V_{\rm esc}$ after \citet{Kokubo:2010}. The
monotonically lowering re-impact velocity is an idealization that
assumes that no interactions with other embryos led to an increase in
the re-impact velocity. A projectile may hit-and-run several times
before finally accreting or being disrupted. The hit-and-run sequence
ended when the collision outcome was partial erosion, partial
accretion, merging, or graze-and-merge. If the hit-and-run sequence
eroded the projectile to less than one third of its original mass, it
was assumed to be debris and neglected. As material was stripped from
the escaping projectile, it was assumed to be derived from the
projectile's mantle, which raised its core mass fraction. In this way,
the core mass fraction of a planet may be enriched after a sequence of
hit-and-run events followed by a merging event.

\subsection{Results}

\begin{table*}[t]
  \caption{\begin{flushleft}Monte Carlo planet growth simulation results for 200 planets using the collision physics model and 
      assuming (group A) no re-impact 
      by a hit-and-run projectile and (group B) with re-impact. Assuming perfect merging only, 
      1455 giant impacts grew 200 planets with final masses 
      up to $2.0 M_{\rm Earth}$. Of these, 50 planets with masses $\ge 0.7M_{\rm Earth}$ grew from 651 giant impacts. 
      For comparison, in the group 3 $N$-body simulations, 52 of 
      161 planets had final masses $\ge 0.7M_{\rm Earth}$. Collision
      material parameters were $c^*=1.9$ and $\bar \mu=0.36$ for hydrodynamic planets. \end{flushleft}}
  \footnotesize
\begin{center}
\begin{tabular}{l || rr | rr || rr | rr  } \hline \hline
   & \multicolumn{4}{c||}{Group A}                                      & \multicolumn{4}{c}{Group B}                                    \\
   & \multicolumn{4}{c||}{No hit-and-run return}                        & \multicolumn{4}{c}{With hit-and-run return}                    \\
   & \multicolumn{4}{c||}{15 Planets $0.7-1.31M_{\rm Earth}$}               & \multicolumn{4}{c}{34 Planets $0.7-1.61M_{\rm Earth}$}       \\
   & \multicolumn{2}{c|}{200 Planets}  & \multicolumn{2}{c||}{$\ge 0.7M_{\rm Earth}$}    & \multicolumn{2}{c|}{200 Planets}  & \multicolumn{2}{c}{$\ge 0.7M_{\rm Earth}$}  \\
  Collision outcome  & $N=1455$ & \% & $N=190$ & \%   & $N=1805$ & \%      & $N=570$ & \%  \\ \hline 
  Super-catastrophic & 3       & $<1$    & 0    & 0                & 5               & $<1$   & 0    & 0     \\
  Partial erosion         & 9      & $<1$    & 1    & $<1$          & 17             & $<1$   & 6   & 1     \\
  Partial accretion     & 489    & 34        & 70   & 37             & 619           & 34      & 194 & 34    \\
  Perfect merging      & 106    & 7         & 17   & 9               & 103            & 6       & 31   & 5     \\
  Graze-and-merge   & 571    & 39      & 56   & 29              & 671           & 37      & 166 & 29    \\
  Hit-and-run (H\&R)   & 277    & 19       & 31   & 16             & 390           & 22      & 139  & 24    \\ \hline \hline
  Special cases          &          &             &      &                    &                  &           &     &       \\
  H\&R with proj. erosion                          & 110        & 8    & 14      & 7   & 185        & 10   & 76    & 13     \\
  $5-10$\% increase in final $f_{\rm core}$ &  43/200 & 22   & 4/15  & 27  & 45/200 & 23   & 9/34  & 26  \\ 
  $>10$\% increase in final $f_{\rm core}$  &  53/200 & 27   & 9/15  & 60  & 71/200 & 36   & 19/34 & 56  \\ \hline \hline
\end{tabular}
\end{center}
\label{tab:mcoutcomes}
\end{table*}

\begin{figure*}[t]
\begin{center}
\includegraphics[width=35pc]{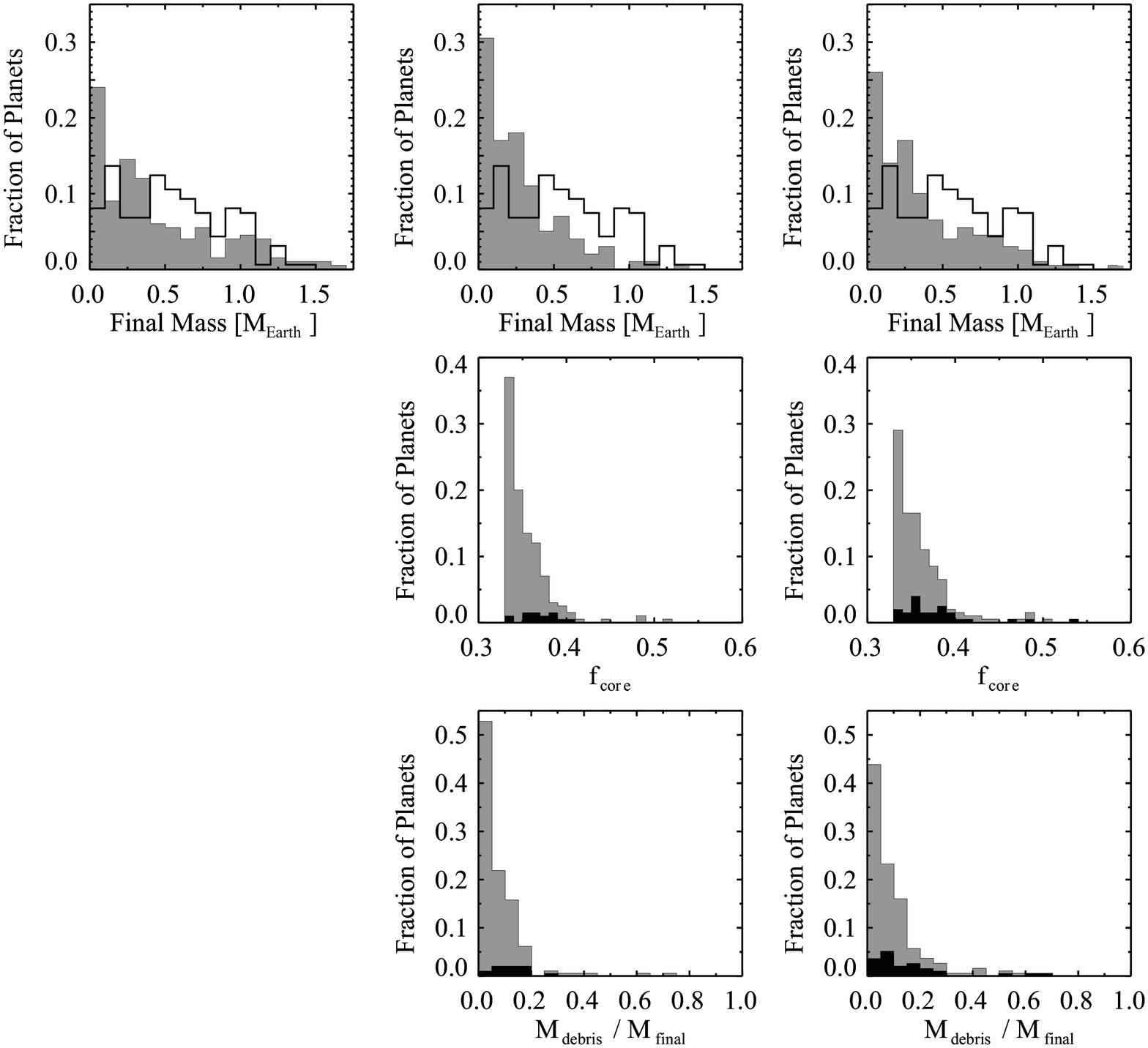}
\end{center}
\caption{Monte Carlo planet growth simulation results. Mass
  distributions, core mass fractions, and debris mass (grey filled
  histograms) for 200 planet growth calculations assuming (A) perfect
  merging, (B,D,F) collision physics model with no hit-and-run return
  collisions, and (C,E,G) collision physics model with re-impact by
  hit-and-run projectile. Black line histogram in A-C is mass
  distribution of planets from group 3 $N$-body simulations. Black
  filled histograms in D-G are for planets with final masses $\ge
  0.7M_{\rm Earth}$.}
\label{fig:montecarlo}
 \end{figure*}

The distribution of final planet masses was calculated for each
simulation set. The random number seed was the same for each group, so
the differences are entirely a results of the collision model
assumptions. The Monte Carlo simulations results are summarized in
Table~\ref{tab:mcoutcomes} and Figure~\ref{fig:montecarlo}. Details of
the simulation parameters in each set is given in appendix figures
\ref{fig:mcsumM}, \ref{fig:mcsumA}, and \ref{fig:mcsumB}. Here, we
focus on the properties of the largest planets with final masses
$>0.7M_{\rm Earth}$. The time between giant impacts was not considered
here because the collision physics model may change the time scale for
planet growth (see \S~\ref{sec:discussion}).

In Figure~\ref{fig:montecarlo}, the mass distribution of final planets
in the perfect merging simulation is similar to the group 3 $N$-body
results, although the Monte Carlo model does not produce the same
number of mid-size planets. The distribution of final planet masses is
significantly smaller when the collision physics model is included and
hit-and-run returns are neglected (group A). In the group 3 set of
$N$-body simulations, 52 of the 161 planets (32\%) had final masses
$>0.7M_{\rm Earth}$. In the perfect merging Monte Carlo simulation, 50
of 200 planets (25\%) reach final masses $>0.7M_{\rm Earth}$; however,
in group A, the number of large planets drops to only 15 (7.5\%). When
hit-and-run return impacts are considered (group B), the final mass
distribution of planets is between perfect merging and group A. In
this case, 34 large planets are produced (17\%).

During collisional growth and fragmentation, material is
preferentially lost from the silicate mantle, thus raising the core
mass fraction (Figure~\ref{fig:montecarlo}D,E). In simulation groups A
and B, the maximum core mass fractions are 0.87 and 0.96,
respectively, in bodies that experienced catastrophic impact
events. Such core-dominated bodies are rare, and most (90-95\%) final
core mass fractions fall in the range of 0.33 to 0.4. In other words,
most of the iron enrichment is within 20\% of the initial value of
$f_{\rm core}$. However, as a group, the largest planets are more
likely to be enriched in core mass fraction compared to smaller
planets. In both group A and B simulations, about 2/3 of the largest
planets have core mass fractions greater than 10\% of the initial
value, compared to about 1/3 of all planets (Table
\ref{tab:mcoutcomes}). The largest planets experience a larger number
of collisions which results in more cumulative erosion of the mantle.

The mass of debris produced during planetary growth by giant impacts
can be significant. While there were planets that suffered only
merging collisions that produced negligible debris (34 planets in
group A and 25 in group B), they all had final masses of less than
$0.38M_{\rm Earth}$ and an average of only 2 giant impacts. For
comparison, the mean number of giant impacts was 7 and 9 for all
planets in groups A and B, respectively.  During the growth of large
planets, debris production averaged 11\% of the final planet mass in
group A and 15\% in group B (Figure~\ref{fig:montecarlo}F,G). The mass
of debris reported in Figure~\ref{fig:montecarlo} only includes debris
from giant impacts; planetesimal collisions would also have
contributed to the debris during planet growth.

In group A, the growth sequence that produced the most debris ($0.28
M_{\rm Earth}$) suffered a penultimate erosive giant impact on a
planet with final mass of only $0.22M_{\rm Earth}$. The most debris
produced from one of the largest planets was 18\% of a $1.31 M_{\rm
  Earth}$ planet.  Notably, there is a case of a $1.2M_{\rm Earth}$
planet in group B that produced $0.83M_{\rm Earth}$ of debris during
its growth. In some cases, the growth sequence includes a step where
the largest remnant is smaller than the initial embryo. Such
destructive sequences occured for 4 planets in group A and 7 planets
in group B.

\begin{figure*}[ht]
\begin{center}
\includegraphics[width=35pc]{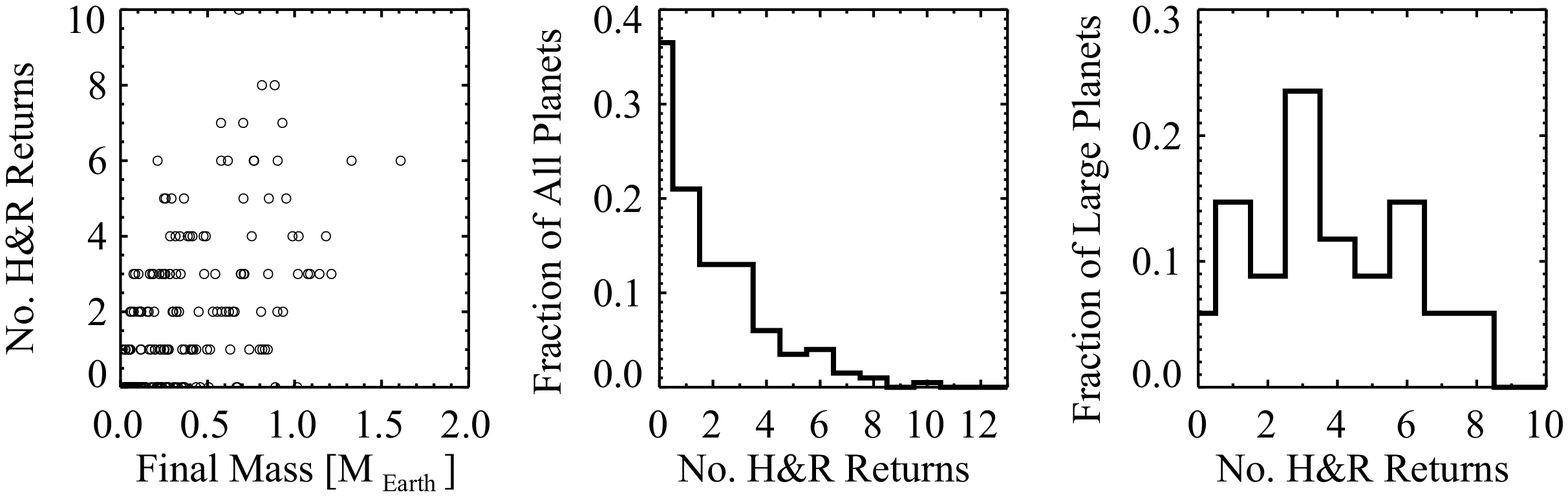}
\end{center}
\caption{Number of return hit-and-run (H\&R) events in the group B
  Monte Carlo planet growth calculations. (A) Number of hit-and-run
  returns vs.\ final planet mass. Histogram of number of return events
  for (B) all 200 planets and (C) 34 planets with final masses $\ge
  0.7M_{\rm Earth}$.}
\label{fig:numhr}
\end{figure*}

An average of 12 giant impacts grew the largest planets in group
A. With the inclusion of hit-and-run return events, an average of 16
giant impacts grew the largest planets in group B.  In group B, the
largest number of giant impacts for any planets was 26 (an ultimately
$0.89M_{\rm Earth}$ planet), in contrast to the maximum of 18 giant
impacts in group A. Hit-and-run return collisions often led to
multiple re-impact events before the final merging or disruption of
the projectile. The number of excess giant impacts in group B is shown
in Figure \ref{fig:numhr}. The mean number of extra collisions on the
largest planets was 4 and ranged from 0 to 8.

\section{Discussion} \label{sec:discussion}

This work demonstrates that the final, stochastic stage of terrestrial
planet formation encompasses a diversity of collision outcomes. All
types of collisions, from super-catastrophic disruption to perfect
merging, are possible. Previous work that assumed perfect merging for
all collisions were not capturing the full complexity of the giant
impact phase. For the expected dynamical conditions for the
terrestrial planets in our early Solar System, the principal
collisions outcomes are approximately evenly split between partial
accretion, hit-and-run, and graze-and-merge.

Because the collision outcome depends on mass ratio, the initial
growth of planets via giant impacts is sensitive to the assumed mass
distribution of planetary embryos. Because different numerical methods
are typically used to model the growth of embryos (the oligarchic
phase) and the end stage of planet formation (the stochastic phase),
the discontinuity in technique has led to significant variation in the
assumed initial number of embryos spread over different radial
distances in different $N$-body studies. For example,
\citet{Raymond:2009} began with 85 to 90 embryos between 0.5 and 4.5
AU and \citet{Obrien:2006} assumed 25 embryos between 0.3 and 4
AU. Both included a population of equal-sized planetesimals. In
contrast, \citet{Kokubo:2006} began with 5 to 30 embryos between 0.5
and 1.5 AU and no planetesimals, and \citet{Kokubo:2010} assumed 16
initial embryos with no planetesimals over the same distance from the
Sun.

In all $N$-body simulations, the initial masses and radial
distributions of embryos are motivated by the isolation mass achieved
at the end of oligarchic growth \citep[e.g.,][]{Kokubo:1998}. However,
the time scale to reach isolation is a function of radial
distance and surface density, whereas the initial state of $N$-body
simulations implies that the isolation masses have been reached
everywhere at the same time. As a result, the number and size
distributions of embryos and planetesimals are not perfectly linked to
earlier stages of planet formation.

To address this problem, a few groups have recently developed hybrid
computational techniques that may link more seamlessly the oligarchic
growth and the stochastic end stage of planet formation
\citep{Levison:2005,Bromley:2006,Bromley:2011,Morishima:2010}. The
current suite of hybrid techniques are based on very different
approaches with different strengths and weaknesses. In this work, we
have demonstrated that the end stage of planet formation involves
significant fragmentation. Because the number of gravitationally
interacting particles is limited in direct $N$-body techniques, hybrid
methods that may include large numbers of smaller bodies (dust to
planetesimals) are required to be able to fully model the giant impact
stage.

\subsection{Fragmentation during the giant impact stage}

Fragmentation is a significant component of the giant impact stage of
terrestrial planet formation. Fragmentation occurs primarily during
partial accretion events and in erosion of the projectile in
hit-and-run events. About 1/3 of giant impacts lead to partial
accretion and about 1/3 are hit-and-run events. Of the hit-and-run
events, about half are energetic enough to erode the projectile (Table
\ref{tab:mcoutcomes}). A small amount of escaping debris is created in
some graze-and-merge events; however, the amount has not been
quantified in our simulations as it is comparable to the resolution
limit (of order 1\% of the mass) \citep{Leinhardt:2012}. In Figure
\ref{fig:collmapequal}, the contours for projectile erosion (blue
dashed line) extend into the graze-and-merge regime for collisions
with a mass ratio more extreme than 1:7. Overall, about half of all
giant impacts create significant (greater than about 1\% of the total
mass) debris.

In contrast, events that lead to erosion of the larger body are rare.
All types of target erosion events (partial erosion to
super-catastrophic disruption) occur about 1\% of the time. Although
infrequent, the conditions required to strip Mercury's mantle by a
single giant impact are included in the range of collision outcomes in
the early Solar System.

To date, numerical simulations that include full fragmentation models
have focused on earlier stages of planet formation, from dust to
planetesimals to oligarchs
\citep[e.g.,][]{Kenyon:1999,Bromley:2011,Leinhardt:2005a,Leinhardt:2009a,
  Chambers:2008, Kobayashi:2010}. In these studies, fragmentation laws
were defined by a size-dependent catastrophic disruption criteria such
as derived by \citet{Benz:1999} (similar to equation
\ref{eqn:qstarred} but including a strength regime for bodies smaller
than about 1 km in radius). The previous fragmentation models have
assumed that the disruption criteria follows pure energy
scaling. Under pure energy scaling, the magnitude of the impact
velocity does not influence the threshold energy for disruption (and
$\bar \mu=2/3$ in equation \ref{eqn:qstarred}). In other words, the
disruption criteria for a particular size target only depended on the
specific energy of the impact $M_{\rm p}V^2_i/M_{\rm t}$. In
\citet{Stewart:2009} and \citet{Leinhardt:2012}, we showed that this
assumption is invalid.

In fact, catastrophic disruption follows nearly pure momentum scaling,
where $\bar \mu=1/3$ \citep{Leinhardt:2012}. As a result, the
disruption threshold is sensitive to the magnitude of the impact
velocity by the factor $V^{*(2-3\bar \mu)}$ in equation
\ref{eqn:qstarred}. Because most numerical and laboratory disruption
experiments have involved projectiles that are much smaller than the
target, most of the published disruption criteria are specifically for
high impact velocities and extreme projectile-to-target mass
ratios. Hence, when the collision involves two bodies that are more
similar in mass, the applied disruption criteria are too high and the
frequency of disruption is underestimated.

\citet{Stewart:2009} compiled disruption data over a wide range of
mass ratios in both the strength and gravity regimes to conclude that
energy scaling was incorrect and the results were closer to momentum
scaling. In \citet{Leinhardt:2012}, we derived a general analytic
model for the disruption criteria in the gravity regime and fit
material parameters for a range of target body types from solid
planetesimals to fluid planets. Future work will tackle catastrophic
disruption in the strength regime.

Now, future models of planet formation have a robust analytic model
for fragmentation during collisions between gravity-dominated bodies.
Our estimate of the magnitude of debris production, on average 15\% of
the mass in the largest planets, is in excellent agreement with
preliminary results from multi-scale $N$-body and hydrocode
calculations. \citet{Genda:2011} calculated the very end stage of
planet formation, beginning with 16 embryos and a total mass of 2.3
$M_{\rm Earth}$, using an $N$-body code. For each collision, the
outcome was calculated by an SPH hydrocode simulation. The largest
remnant or the two hit-and-run bodies were re-inserted into the
$N$-body calculation and the debris was neglected. The cumulative mass
of the fragments was about 22\% of the total mass ($0.48 M_{\rm
  Earth}$). Such multi-scale calculations are extremely
computationally expensive, and the analytic model will allow for more
and more detailed investigations of planet formation.

The need to track many small fragments is a significant challenge in
models of the end stage of planet formation. The dynamical
interactions between the protoplanets must be calculated by direct
$N$-body techniques, but the smaller fragments will need to be treated
statistically. As a result, detailed models of the giant impact stage
must adopt new methods, such as the hybrid codes mentioned
above. Hybrid models are needed to be able to calculate what fraction
of the debris is reaccreted onto the final planets and what fraction
is removed (e.g., via Poynting-Robertson drag). For fragments ground
down below about 1 km in size, new disruption criteria are still
needed in the strength regime that fully account for material
properties, impact angle, mass ratio, and impact velocity.

\subsection{Time scale of planet formation}

The influence of more realistic collision physics on the time scale of
terrestrial planet formation is difficult to predict because of
competing factors. On one hand, the growth rate of planets is slower
when outcomes other than perfect merging are included. On the other
hand, the debris produced by collisions can influence the overall
dynamics of the embryos and planetesimals (e.g., via dynamical
friction, which could lead to more lower velocity collisions and more
frequent merging outcomes).

The time scale for the end stage of planet formation was investigated
in a recent study that included two collision
outcomes. \citet{Kokubo:2010} conducted $N$-body simulations with a
collision model that allowed for either perfect merging or an ideal
hit-and-run event using their empirical boundary presented in equation
\ref{eqn:vhr} (which they applied at all impact angles). In a
hit-and-run collision, neither body lost mass (there was no
fragmentation) but the relative velocities of the bodies decreased. In
simulations that began with 16 equal-mass embryos between 0.5 and 1.5
AU, they found that about half of the collisions were hit-and-run
events. After a hit-and-run encounter, the reduced relative velocities
led to a high probability of merging on the subsequent encounter. As a
result, the time scale for planet growth was essentially the same as
in simulations with the same starting conditions that assumed perfect
merging.

The magnitude of any changes to planet growth time scales from
different collision outcomes is sensitive to the initial conditions in
the simulation. The simulations by \citet{Kokubo:2010} began with few
relatively large embryos ($0.15M_{\Earth}$) and no planetesimals. With
this initial distribution of embryos, all collisions were between
comparable mass bodies ($M_{\rm p}/M_{\rm t}>0.1$), which are the most
likely to be graze-and-merge or hit-and-run events (Figure
\ref{fig:collmapequal}). Since the hit-and-run bodies are likely to
recollide on time scales comparable to the orbital period, the overall
effect of hit-and-run events on this very end stage of planet growth
was found to be negligible. If fragmentation were included or if the
sizes of the bodies were more diverse (e.g., with the inclusion of
planetesimals or smaller initial embryos), more realistic collision
outcomes would have a larger effect on planet growth. As mentioned
above, using the same initial conditions in their multi-scale
calculations, \citet{Genda:2011} found that fragmentation was a
significant process.

Although fragmentation makes planet growth from an individual
collision less efficient, other effects from the debris may lead to
faster planet growth overall or faster stages of planet growth. For
example, during oligarchic growth, \citet{Chambers:2008} found that
fragmentation led to faster growth of embryos because smaller
fragments are more easily captured. However, fragmentation also
decreased the surface density of solids in the disk (which limits the
embryo's final mass) because fragments were lost more quickly by drag
processes as they were ground down in size. During the giant impact
stage, \citet{Obrien:2006} found that the strong dynamical friction
from 1000 planetesimals led to overall faster planet growth compared
to studies without any planetesimals. However, their study assumed
perfect merging for all collision outcomes and non-interacting
planetesimals. If dynamical friction becomes very large, a gap could
form in the planetesimal disk around an embryo, which would
effectively halt the growth of that embryo. Hence, the role of
fragmentation on planet growth time scales is not independent of other
processes acting at the same time. Ultimately, because fragmentation
is a critical process that feeds back into the dynamics of planet
growth, new simulations that include both embryos and a fully
interacting population of small bodies are needed to investigate how
more realistic collision outcomes affect formation time scales.

\subsection{Composition and chemistry of planets}

We found that the core-to-mantle mass fraction increases during the
growth of planets via fragmentation during collisions between
differentiated embryos. Loss of the silicate mantle primarily occurred
during partial accretion and erosion of the projectile in hit-and-run
events.  From our group B Monte Carlo calculations, the mean increase
in the core mass fraction was 15\% for the 34 largest planets (Figure
\ref{fig:montecarlo}E, black histogram). The magnitude of the core
fraction increase is potentially observable in the study of the
chemical composition of planets and early Solar System materials.

\citet{Oneill:2008} argue that the Earth's bulk iron to magnesium
(Fe/Mg) ratio is significantly larger than the solar ratio. They
propose that the Earth lost a portion of its silicate mantle during
the giant impact stage, which raised the Fe/Mg ratio of the final
planet compared to the more primitive (closer to nebular composition)
materials that formed the planetary embryos. \citet{Oneill:2008}
estimate that the whole-Earth Fe/Mg mass ratio is $2.1\pm0.1$. The
Fe/Mg value for primitive materials is not known precisely. The solar
photosphere value, $1.87\pm0.4$ \citep{Asplund:2005}, is too poorly
constrained to be useful for such detailed comparisons. The Fe/Mg
ratio for the solar wind will be constrained by Genesis mission.
Early results suggested a lower value than the solar photosphere
\citep[$1.61\pm0.23$,][]{Jurewicz:2011}; however, final data
calibration is still in progress (A.\ Jurewicz, pers.\ comm.). The
Fe/Mg ratio for carbonaceous chondrites, the most primitive type of
meteorite, is $1.92\pm0.08$ \citep{Palme:2005}. The available data
suggest that the Earth is enriched in Fe/Mg compared to solar
composition by approximately 10\%.

Our Monte Carlo calculations of the cumulative effects of realistic
collision outcomes are in excellent agreement with the collisional
erosion idea proposed by \citet{Oneill:2008}. Although our planet
growth models are not detailed enough to track individual elements, we
use the core-to-mantle mass ratio as a proxy for the Fe/Mg
ratio. About half of the largest planets in the group B calculations
have core mass fraction enrichments of 10-30\%. The mean is likely to
be slightly lower than this range because some of the mantle fragments
will be reaccreted. Assuming that the reaccretion process is less than
100\% efficient, the largest terrestrial planets have a larger core
mass fraction compared to the initial embryo composition as a result
of fragmentation during giant impacts.

Bulk elemental ratios are difficult to derive for a planet, and more
detailed information can be derived from examination of isotopic
systems. The hafnium-tungsten (Hf-W) system, with a half life of 9
Myr, is a major constraint on the timing of planet formation. Both
elements are refractory (high condensation temperatures); Hf is
lithophile and is retained in the silicate mantle, while W is
moderately siderophile and prefers the iron core. With assumptions
about the extent of chemical equilibration between the metals and
silicates within the growing planet, measurements of the present-day
ratios of Hf and W isotopes constrain the time scale for planet
formation, including the timing of the proposed Moon-forming impact
\citep{Jacobsen:2005,Kleine:2009}.

Using 15 planets from the $N$-body simulations by \citet{Obrien:2006}
(Group 1 in Table \ref{tab:outcomes}), \citet{Nimmo:2010} calculated
the evolution of the Hf-W system during planet growth in the giant
impact stage and compared the results to observed values for Earth,
the Moon, and Mars. The study varied the equilibration factor between
two idealized end members: no equilibration, where the core from the
projectile merges with the target core without any equilibration with
the mantle, to perfect equilibration, where the projectile core
completely equilibrates with the mantle. The equilibration factor was
held constant during the entire growth of a planet; in reality, the
extent of equilibration depends on the physics of material mixing
during giant impacts, which is still rather poorly understood
\citep[e.g.,][]{Dahl:2010}. \citet{Nimmo:2010} found that Earth-like
Hf-W ratios could be generated if the iron cores partially
equilibrated with the mantle; however, they could not simultaneously
match both the Earth and the Moon. They suggest that either that the
Earth and Moon equilibrated after the giant impact
\citep{Pahlevan:2007} or that the time scales for planet growth are
too short in the $N$-body simulations.

With more realistic collision physics in planet formation
simulations, calculations of the evolution of the Hf-W system will be
more robust. In particular, the collision model estimates the fraction
of core and mantle incorporated during partial accretion of the
projectile and following hit-and-run events with projectile
erosion. Improving the time scale for planet growth, as discussed
above, is essential for interpretation of the Hf-W system. Finally,
improved physical models for mixing of metals and silicates during
giant impacts are still needed.

Fragmentation during planet formation may also affect the final
volatile content. Several studies have investigated possible sources
of water in the Earth by assuming that the water content in condensed
material varies monotonically with distance from the Sun. The initial
locations of material that accretes into terrestrial planets have been
used to investigate the formation of Earth-like planets under the
assumption of perfect merging
\citep[e.g.,][]{Morbidelli:2000,Raymond:2004,Raymond:2007}.  The
accretion of water to the Earth may now be investigated in greater
detail.

The incorporation of volatiles in planets not only depends on the
source location of the incoming material but also the specific
collision scenario. \citet{Asphaug:2010} proposed that the volatile
content of planetary bodies may be affected by hit-and-run
events. Because erosion of the projectile is common in hit-and-run
events, the projectile may be stripped of volatile-rich outer layers
(including an atmosphere). If the hit-and-run projectile is later
incorporated into a growing planet, the planet would be depleted in
both volatiles and mantle material compared to the initial embryos.

The process of compositional changes via fragmentation is not
restricted to the inner Solar System.  The dwarf planets in the outer
Solar System have higher bulk densities than smaller Kuiper Belt
objects \citep{Brown:2008,Fraser:2010}. The bulk densities of dwarf
planets may have increased during collisional growth by preferential
stripping of the icy mantles from projectiles, possibly by partial
accretion of planetesimals during runaway growth in addition to the
limited number of giant impacts experienced by dwarf planets.

\subsection{The graze-and-merge regime}

The boundary between the graze-and-merge and hit-and-run regimes is
not precisely known. In this work, we used the boundary defined from a
single set of calculations for differentiated iron-silicate planets
\citep{Kokubo:2010, Genda:2012}. The boundary is slightly different
in the few other studies that have focused on this regime. In Figure
\ref{fig:collmapequal}, the results from different studies are plotted
as symbols for comparison to the \citet{Kokubo:2010} graze-and-merge
boundary. 

\citet{Leinhardt:2000} calculated the outcome of collisions between
equal-mass rubble piles (black symbols\footnote{These data are derived
  from Table 1 in \citet{Leinhardt:2000}. The graze-and-merge regime
  is identified as when the accreting mass is equal to the projectile
  mass (e.g., cases where the graze-and-merge was a long duration
  event); some of the points plotted as perfect merging outcomes may
  have gone through a fast graze-and-merge event.}). The collision
velocities were subsonic between the 1-km radius bodies, so the study
utilized the pkdgrav $N$-body code rather than a shock hydrocode. The
results show that grazing outcomes extend to impact parameters
slightly below $b_{\rm crit}$. For impact angles between about 30 and
50 degrees, the transition to hit-and-run is in good agreement with
\citep{Kokubo:2010}. However, at higher impact angles ($64^{\circ}$),
the rubble pile collisions transitioned to hit-and-run at much lower
impact velocities. Using the same code, \citet{Leinhardt:2010} modeled
collisions onto $\sim700$ km radius bodies with equal mass and half
mass projectiles (blue symbols in Figure \ref{fig:collmapequal}A and
B). The results for equal mass bodies support the same boundary as
\citep{Kokubo:2010} for the 30 to $40^{\circ}$ impacts. However, the
simulations for a mass ratio of 1:2 show a much faster transition to
hit-and-run at an impact angle of $53^{\circ}$.

The different boundaries between graze-and-merge and hit-and-run may
be primarily attributed to the relatively low resolution for all three
studies. The studies were focused on other aspects of collisions and
not to designed to resolve the interaction of very thin layers of
material at high impact angles. In addition, small differences in
dissipation of energy and momentum are likely to influence the
transition to hit-and-run at high impact angles. If the
graze-and-merge regime is smaller than considered here, the collisions
would increase the fraction of hit-and-run events.  

The graze-and-merge regime is believed to have left a strong mark in
our Solar System. The formation of Earth's Moon \citep{Canup:2004},
the Pluto system \citep{Canup:2005}, and the Haumea system
\citep{Leinhardt:2010} are all attributed to graze-and-merge regime
events (Figure \ref{fig:collmapequal}).  More work is needed to
understand the details of moon formation in the graze-and-merge regime
and the factors that control the transition between the
graze-and-merge and hit-and-run regimes.

\section{Conclusions} \label{sec:conclusions}

Using a new analytic collision physics model \citep{Leinhardt:2012},
we have investigated the range of collision outcomes during the
stochastic end stage of planet formation. For the dynamical conditions
expected in our early Solar System, the outcome of giant impacts span
all possible regimes: hit-and-run, merging, partial accretion, partial
erosion, and catastrophic disruption.  Fragmentation during giant
impacts is significant. During the formation of planets larger than
$0.7M_{\rm Earth}$, the total mass of debris is about 15\% of the
final planet mass.  Fragmentation occurs primarily by erosion of the
smaller body in partial accretion and hit-and-run events. Future
simulations of the end stage of planet formation will need to utilize
hybrid techniques that are able to track both massive planets and a
large population of smaller bodies.  Assuming that fragments are not
completely reaccreted, growth via giant impacts creates final planets
are that depleted in volatiles and mantle material compared to the
initial planetary embryos.

{\it Acknowledgements.} We thank D. O'Brien and S. Raymond for the
$N$-body simulation data and valuable discussions. STS is supported by
NASA grant \# NNX09AP27G, ZML by a STFC Advanced Fellowship.


\clearpage
\clearpage

{\bf APPENDIX FIGURES}
\setcounter{figure}{0}
\renewcommand{\thefigure}{A.\arabic{figure}}
\setcounter{table}{0}
\renewcommand{\thetable}{A.\arabic{table}}

\begin{figure}[ht]
\includegraphics[width=30pc]{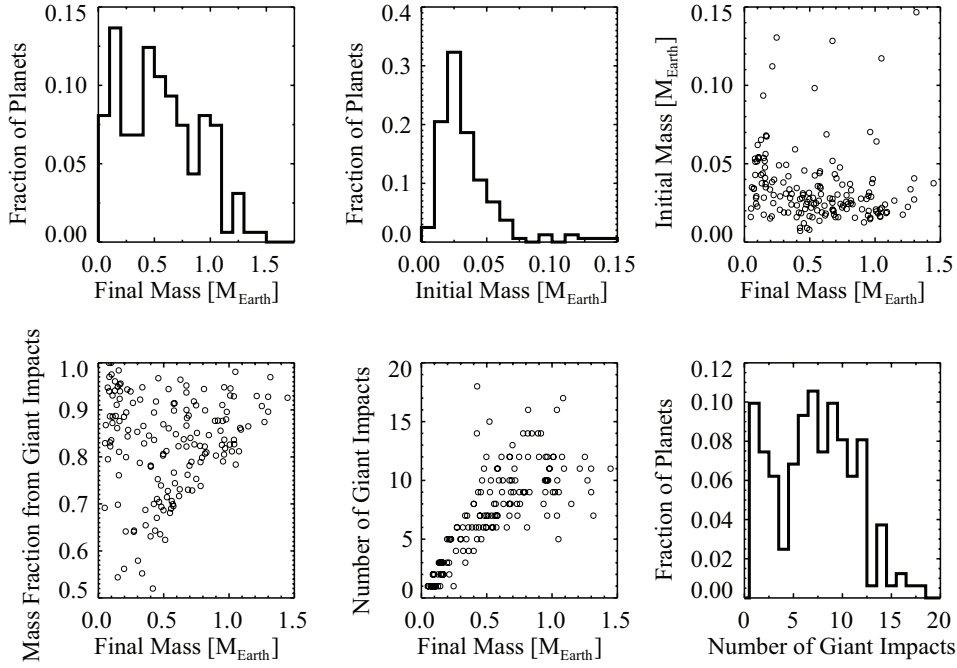}
\caption{Distribution of parameters from group 3 $N$-body
  simulations. Distributions for initial mass, number of giant
  impacts, and mass contribution from planetesimals used in Monte
  Carlo planet growth simulations. }
\label{fig:group3}
 \end{figure}
 \clearpage

\begin{figure}[ht]
\includegraphics[width=35pc]{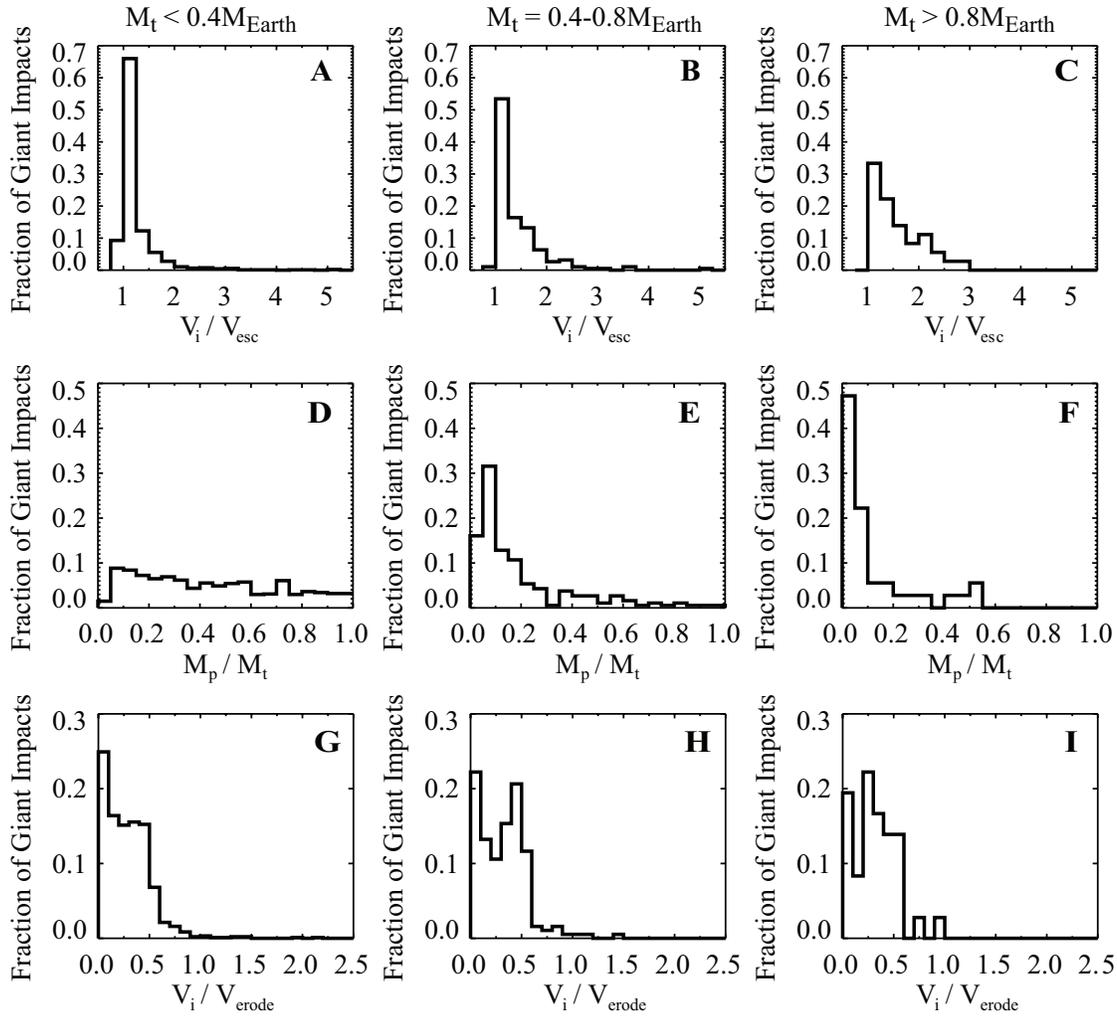}
\caption{Distribution of impact velocities (A-C), projectile-to-target
  mass ratios (D-F), and ratio of impact velocity to velocity needed
  for onset of target erosion ($V_{\rm erode}$) (G-I) for all giant
  impacts in group 3 for $M_{\rm t}<0.4M_{\rm Earth}$ (left column),
  $0.4M_{\rm Earth}<M_{\rm t}<0.8M_{\rm Earth}$ (middle column), and
  $M_{\rm t}>0.8M_{\rm Earth}$ (right column). Target mass-dependent
  values for impact velocity and projectile-to-target mass ratios used
  in Monte Carlo planet growth simulations.}
\label{fig:group3vels}
 \end{figure}
 \clearpage

\begin{figure}[ht]
\includegraphics[width=35pc]{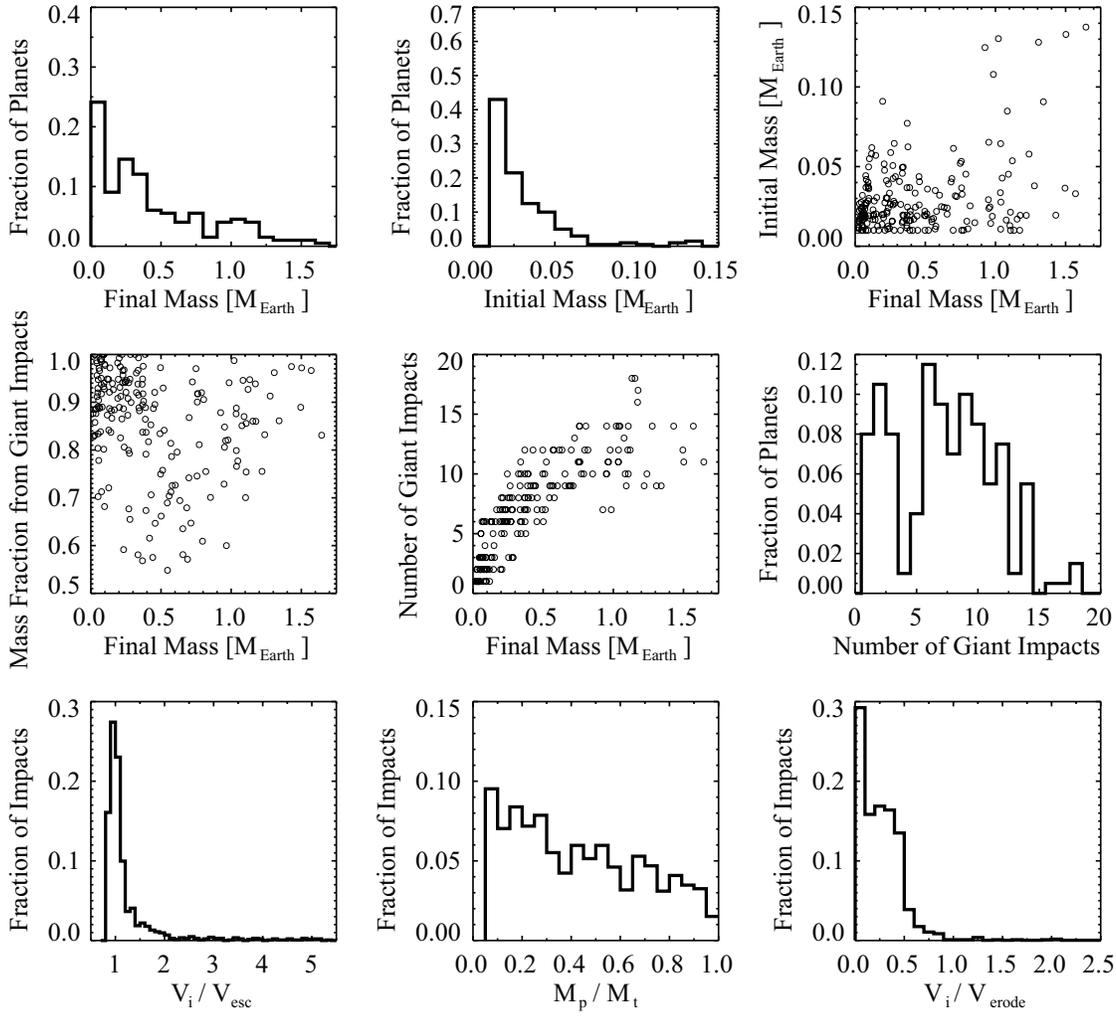}
\caption{Distribution of parameters for 200 Monte Carlo planet growth simulations with perfect merging.}
\label{fig:mcsumM}
 \end{figure}
 \clearpage

\begin{figure}[ht]
\includegraphics[width=35pc]{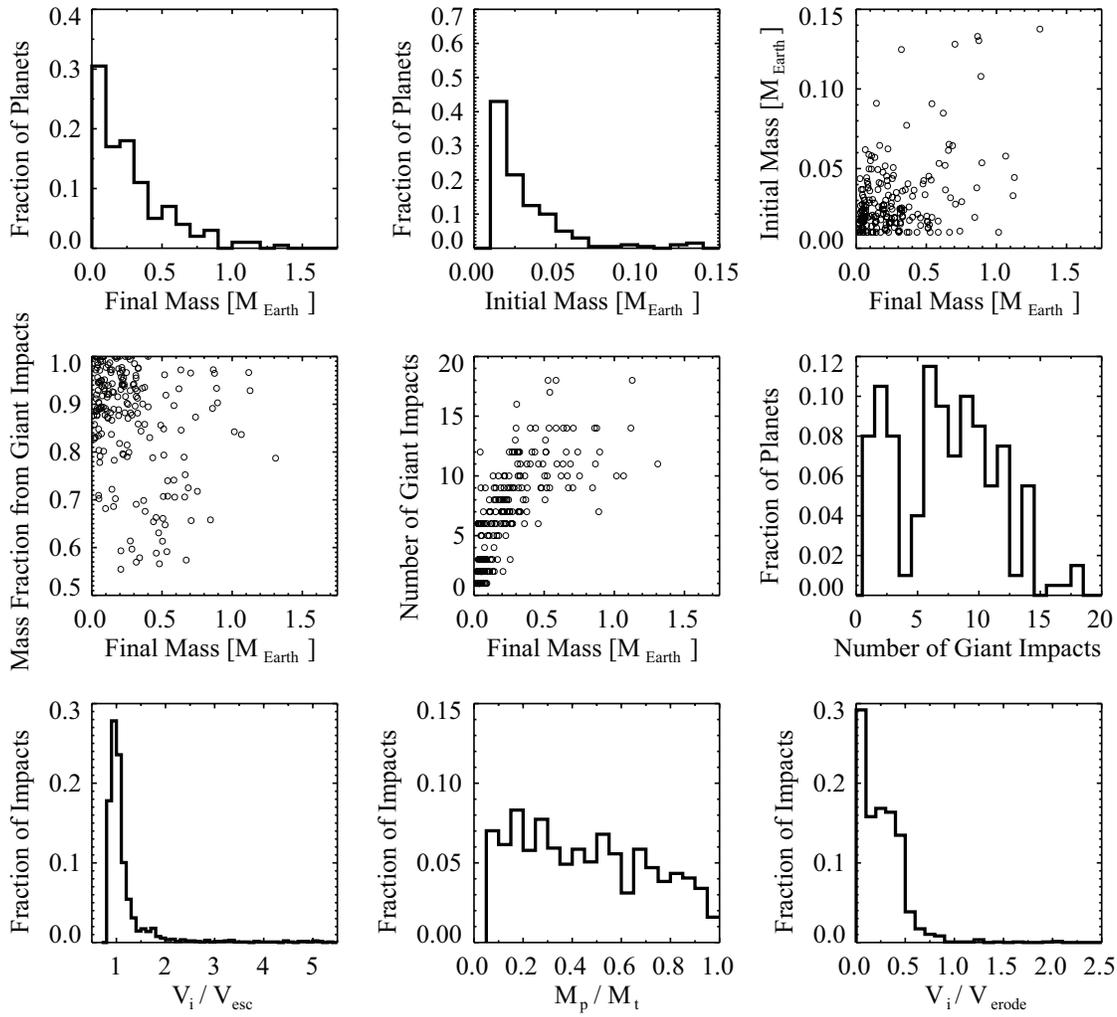}
\caption{Distribution of parameters for 200 Monte Carlo planet growth simulations with collision physics model and assuming no re-impact by a hit-and-run projectile (group A).}
\label{fig:mcsumA}
 \end{figure}
 \clearpage

\begin{figure}[ht]
\includegraphics[width=35pc]{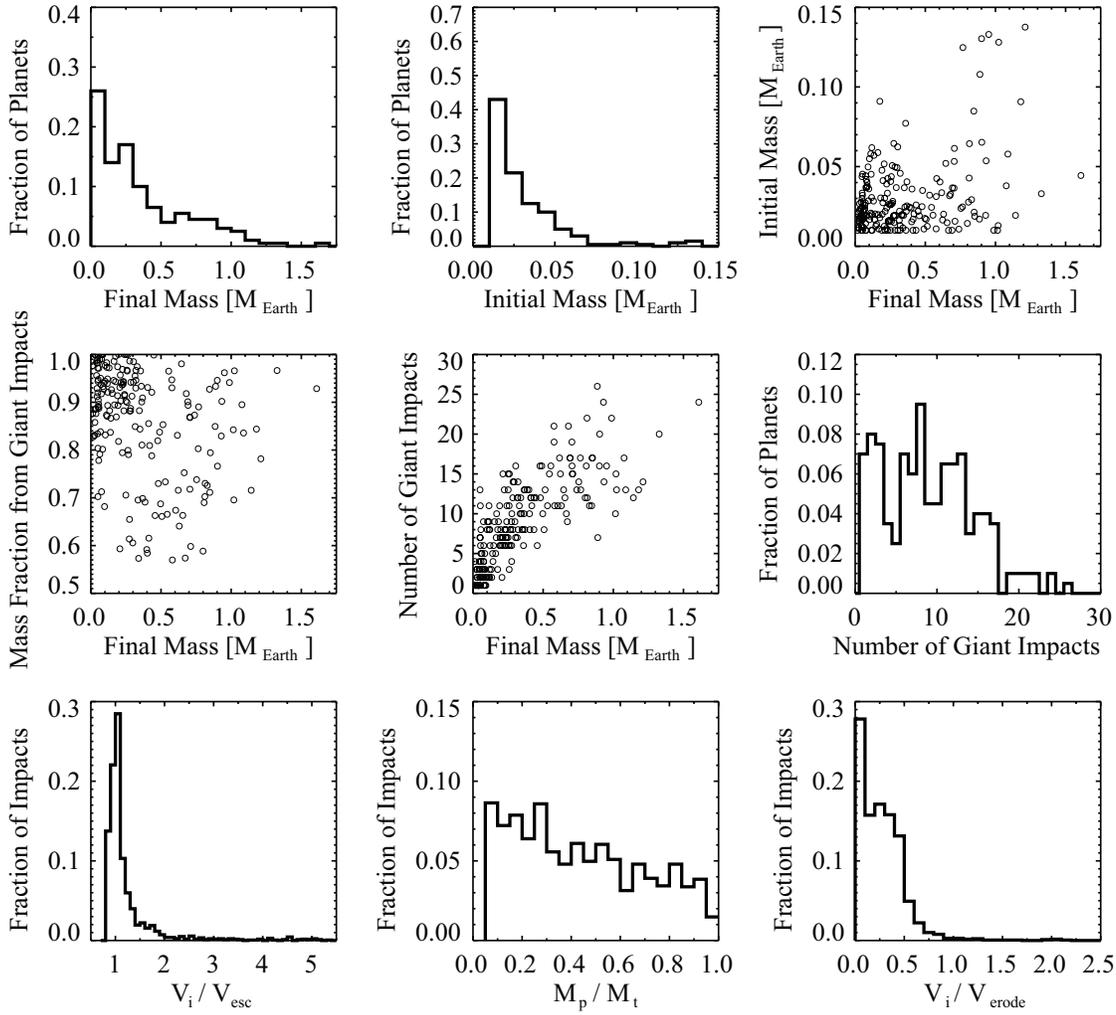}
\caption{Distribution of parameters for 200 Monte Carlo planet growth
  simulation with collision physics model with re-impact by a
  hit-and-run projectile (group B). Note the increase in number of
  giant impacts compared to group A.}
\label{fig:mcsumB}
 \end{figure}

\end{document}